\documentclass[12pt]{article}
\usepackage{graphicx}
\renewcommand{\baselinestretch}{1.0}

\newcommand{\be}{\begin{equation}}
\newcommand{\ee}{\end{equation}}

\def\aprge{\buildrel > \over {_{\sim}}}
\newcommand{\bea}{\begin{eqnarray}}
\newcommand{\eea}{\end{eqnarray}}
\begin{document}
\topmargin 0pt
\oddsidemargin=-0.4truecm
\evensidemargin=-0.4truecm
\renewcommand{\thefootnote}{\fnsymbol{footnote}}

\newpage
\setcounter{page}{1}
\begin{titlepage}
\vspace*{-2.0cm}
\begin{flushright}
\vspace*{-0.2cm}

\end{flushright}
\vspace*{0.5cm}

\begin{center}
{\Large \bf Solar neutrino spectrum, sterile neutrinos and
 additional radiation in the Universe}
\vspace{0.5cm}

{P. C. de Holanda$^{1}$ and  A. Yu. Smirnov$^{2}$\\

\vspace*{0.2cm}
{\em (1) Instituto de F\'\i sica Gleb Wataghin - UNICAMP,
13083-970 Campinas SP, Brazil}\\
{\em (2) The Abdus Salam International Centre for Theoretical Physics,
I-34100 Trieste, Italy }
}
\end{center}

\vskip 1cm

\begin{abstract}

Recent results from the SNO, Super-Kamiokande and Borexino experiments  
do not show the expected upturn of the energy spectrum of events 
(the ratio $R \equiv N_{obs}/N_{SSM}$) at low energies. 
At the same time, cosmological observations  testify for possible  
existence of additional relativistic degrees of freedom in the early Universe:  
$\Delta N_{eff} = 1 - 2$. These facts strengthen the case  
of very light sterile neutrino, $\nu_s$, 
with  $\Delta m^2_{01} \sim (0.7 - 2) \cdot 10^{-5}$ eV$^2$, 
which mixes weakly with the active neutrinos. 
The $\nu_s$ mixing in the mass eigenstate $\nu_1$ characterized by  
$\sin^2 2\alpha \sim 10^{-3}$ can explain an absence of the upturn. 
The mixing of $\nu_s$ in the eigenstate $\nu_3$ with  $\sin^2 \beta \sim 0.1$
leads to production of $\nu_s$ via oscillations in the  Universe 
and to additional contribution $\Delta N_{eff} \approx 0.7 - 1$ 
before the big bang nucleosynthesis and later. 
Such a mixing can be tested in forthcoming experiments with 
the atmospheric neutrinos as well as in future 
accelerator long baseline experiments. It has substantial impact on 
conversion of the supernova neutrinos. 

\end{abstract}

\end{titlepage}
\renewcommand{\thefootnote}{\arabic{footnote}}
\setcounter{footnote}{0}
\renewcommand{\baselinestretch}{0.9}

\section{Introduction}

The large mixing angle (LMA) MSW solution \cite{w1,ms}
has been established as the solution of the 
solar neutrino problem 
\cite{hom}, 
\cite{Abdurashitov:2009tn}, 
\cite{Altmann:2005ix}, 
\cite{Hosaka:2005um}, 
\cite{Aharmim:2005gt}, 
\cite{Aharmim:2007nv}, 
\cite{Aharmim:2008kc}, 
\cite{Arpesella:2008mt}. 
In assumption of the CPT conservation KamLAND confirms this result  
\cite{:2008ee}, \cite{Gando:2010aa}.
One of the main goals of further precision measurements of the 
solar neutrino fluxes is to search for possible deviations from 
the LMA predictions  which would indicate 
new physics beyond the Standard Model with three mixed neutrinos. 
In particular, new physics can show up  
at the neutrino  energies $E = (1 - 7)$ MeV, {\it i.e.} in the transition 
region between the matter dominated conversion and 
vacuum oscillations. Here direct measurements of the spectrum are absent or 
inprecise and possible deviations from the LMA predictions can be relatively large.  

Some time ago in attempt to explain the 
low (about $2\sigma$) rate in the  Homestake experiment~\cite{hom} 
in comparison to the LMA expectation as well as the absence of clear 
low energy upturn of the spectra of events at SuperKamiokande  and SNO 
we have proposed a scenario with light 
sterile neutrino, $\nu_s$, which mixes weakly with active neutrinos \cite{pedroS}. 
Conversion of $\nu_e$ to $\nu_s$ driven by the mass squared difference  
$\Delta m^2_{01} \sim (0.2 - 2) \cdot 10^{-5}$ eV$^2$  
and mixing in the mass state $\nu_1$,  
$\sin^2 2\alpha \sim 10^{-3}$, leads 
to appearance of a dip in the $\nu_e - \nu_e$ survival 
probability in the range  (0.5 - 7) MeV which explains the data.  

After  publication  \cite{pedroS} several new experimental results have appeared which further support our proposal: 

\begin{itemize} 

\item 

Measurements of the solar neutrino spectrum by SuperKamiokande-III \cite{Abe:2010hy} with lower threshold still do not show  the upturn. 

\item 

The SNO LETA analysis \cite{Aharmim:2009gd} gives even 
turn down of the spectrum in the two lowest energy bins.   

\item 

The Borexino measurements of the boron neutrino spectrum also hint some tendency of the spectral turn down \cite{Bellini:2008mr}.

\end{itemize}

Although separately these results are not statistically significant,  
being  combined  they can be considered an evidence of some new 
sub-leading effect.  

At the same time, the cosmological observations indicate possible presence of 
additional radiation in the Universe in the epoch of last photon scattering. 
This is quantified by the effective number of neutrino species,  
$N_{eff}$, which is bigger than 3. Combined analysis of WMAP-7,  
measurements of BAO (Baryon Acoustic 
Oscillations) and new value of the Hubble constant $H_0$) gives  
$N_{eff} = 4.34^{+ 0.86}_{- 0.88}$ \cite{Komatsu:2010fb}. 
WMAP-7 and  Atacama Cosmology Telescope data lead to  
$N_{eff} = 5.3 \pm 1.3$  (68 \% C.L.) \cite{Dunkley:2010ge}. 
In the independent  
analysis~\cite{Hamann:2010bk} of these data the number of very light  
sterile neutrinos $\Delta N_{eff} = (0.02 - 2.2)$ (68 \% C.L.) has been obtained. 
All this confirms the earlier finding based on the WMAP-3 data: 
$N_{eff} = 5.3^{+ 0.4}_{- 0.6}{}^{+ 2.1}_{- 1.7}{}^{+ 3.8}_{-2.5}$~\cite{Seljak:2006bg}. 

These results do not contradict the recent Big Bang Nucleosynthesis (BBN)
bounds $N_{eff} = 3.68^{+ 0.80}_{- 0.70}$ \cite{Izotov:2010ca} 
(see  discussion in \cite{BBN} and theoretical considerations in \cite{BBNint}). 
Hence an additional radiation can be produced before  the BBN epoch. 

In this connection we revisit our proposal of very light sterile neutrinos. 
We show that mixing of this  neutrino in mass states $\nu_1$ 
or/and $\nu_2$  can consistently improve 
description of the solar spectral data. We introduce  
mixing of this neutrino in the  
mass eigenstate $\nu_3$  which allows $\nu_s$ to be produced in the 
Early Universe with nearly equilibrium concentration, so that 
$\Delta N_{eff} \approx 1$.

The paper is organized as follows. In sect. 2 we 
consider properties of the  $\nu_e$ conversion 
in the presence of $\nu_s-$mixing in the Sun generalizing our analysis in \cite{pedroS}.  
New feature, wiggles'' in the survival probability, is described  
which appear for relatively large $\Delta m^2_{10}$  at the $E > 5$ MeV. 
In sect. 3 we obtain 
bounds on the $\nu_s$ parameters from the Borexino measurements of the 
$Be-$neutrino flux. Spectra of the solar neutrino events 
have been computed for different experiments 
and confronted  with the data. 
In sect. 4. the  mixing of $\nu_s$ in $\nu_3$ is introduced and 
phenomenological consequences of this mixing are studied,  
in particular, generation of $\nu_s$ in the Early Universe.   
The conclusion is given in sect. 5. In appendix we 
give some details of appearance of  
the wiggles in the survival probability.  

\section{Sterile neutrino and conversion probabilities}

\subsection{Generalities}

Let us consider the system of 4  neutrinos $\nu_f = (\nu_s, \nu_e, \nu_\mu, \nu_\tau)$
mixed in the mass eigenstates $\nu_i$, $i = 0, 1, 2, 3$. 
The sterile neutrino, $\nu_s$, is mainly present in the mass eigenstate $\nu_0$
with mass $m_0$. It mixes weakly with active
neutrinos and this mixing can be treated as small perturbation 
of the standard LMA structure. 
  
Coherence of all mass eigenstates is lost on
the way to the Earth. Therefore  the $\nu_e$-survival probability
at the surface of the Earth can be written as 
\be   
P_{ee} = \sum_i |A_{ei}^S|^2 |U_{ei}|^2~,
\label{eq:gen1}
\ee
where $A_{ei}^S$ is the amplitude 
of the $\nu_e \rightarrow \nu_i$ transition inside the Sun and  
$U_{ei} \equiv \langle \nu_e| \nu_i \rangle$ 
is the element of the mixing matrix in vacuum. 
The quantities in eq. (\ref{eq:gen1}) satisfy the normalization conditions: 
$\sum_i |U_{ei}|^2 = 1$ and 
$$ 
\sum_i |A_{e i}^S| = 1.  
$$
During nights the solar neutrinos oscillate in the matter of the Earth.  
In this case  $U_{ei}$ in eq. (\ref{eq:gen1}) should be substituted by 
the $\nu_i \rightarrow \nu_e$ oscillation probabilities  
inside the Earth, $U_{ie} \rightarrow  A_{ie}^E$, so that   
\be
P_{ee} = \sum_i |A_{ei}^S|^2 |A_{ie}^E|^2~. 
\label{eq:earth}
\ee

In the production point  the electron neutrino state 
can be represented in terms of the eigenstates in matter, $\nu_{im}$, as 
$$
\nu_e = \sum_i U_{ei}^m \nu_{im},  ~~~(i = 0, 1, 2, 3),  
$$
where $U_{ei}^m$ is the mixing matrix element in matter in the production region. 
We denote by $\lambda_i$  the eigenvalues which correspond to the 
eigenstates $\nu_{im}$.  
Introducing  $A_{ji}$ --  the amplitudes of $\nu_i^m \rightarrow \nu_j$ 
transitions inside the Sun
we can write
\be
A_{ei}^S = \sum_j U_{ej}^{m} A_{ji}.  
\label{eq:amplij}
\ee
Insertion this expression into (\ref{eq:earth}) gives   
\be
P_{ee} = \sum_i |\sum_j U_{ej}^{m} A_{ji}|^2 |A_{ie}^E|^2~. 
\label{eq:total}
\ee
In the adiabatic case  $A_{ij} = \delta_{ij}$, so that 
\be
P_{ee} = \sum_i |U_{ei}^{m}|^2 |A_{ie}^E|^2~. 
\label{eq:totalad}
\ee


\begin{figure}[ht]
\begin{center}
\vskip 1cm
\includegraphics[width=13cm]{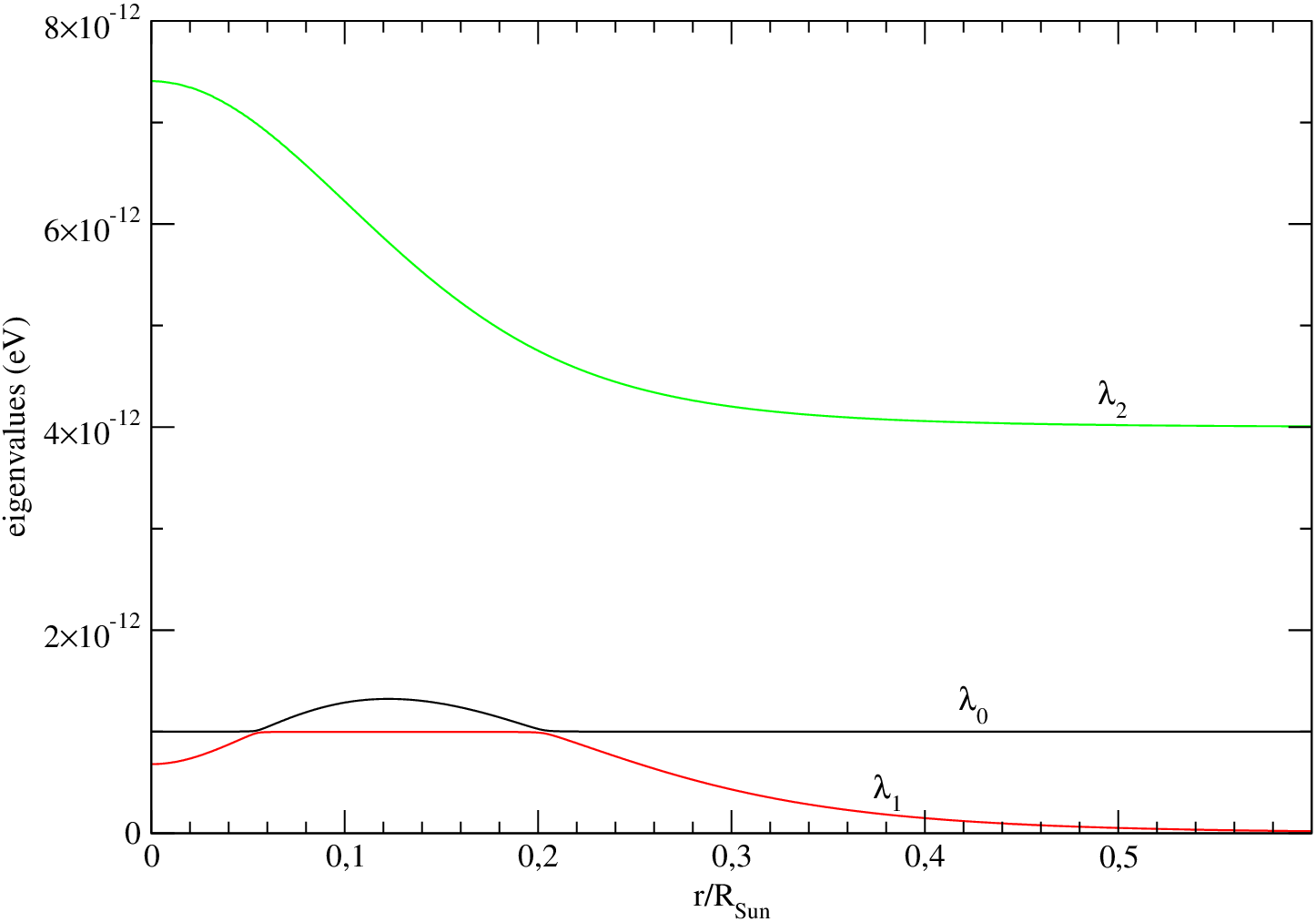}
\caption{The level crossing scheme for the neutrino energy 
$E = 10$ MeV. 
Dependence of the eigenvalues of total Hamiltonian 
in matter, $\lambda_i$, on distance from the center of the Sun. 
The LMA  neutrino oscillation parameters are taken as  
$\Delta m^2_{21} = 8\times 10^{-5}$ eV$^2$ and $\tan^2\theta_{12} =0.44$, 
while sterile neutrino parameters are $R_\Delta = 0.25$ and $\sin^22\alpha=10^{-3}$.}
\label{fig:eigenvalues}
\end{center}
\end{figure}

For low energies we are interested in, the Earth matter effect is small and 
can be neglected in the first approximation.

In what follows we will introduce mixing of the sterile neutrino 
in different mass eigenstates.  
In computations of effects for solar neutrinos
we  neglect the 1-3 mixing 
and therefore consider the mixing of only three  
flavor states $\nu_f^{(3)} = (\nu_s, \nu_e, \nu_a)$
($\nu_a$ is the mixture of $\nu_\mu$ and $\nu_\tau$).  
The mixing matrix which connects these states 
with the mass eigenstates, $\nu_{mass} = (\nu_0, \nu_1, \nu_2)$, can be parametrized as 
\be
U^{(3)} =  U_{\theta} U_{\alpha}, 
\label{eq:3mix}
\ee
where $U_{\alpha}$ is the matrix which mixes $\nu_s$ in  $\nu_1$ or/and $\nu_2$
and $U_{\theta} \equiv U_{12}(\theta_{12})$ is the standard LMA mixing (the rotation 
by the angle $\theta_{12}$ in the $\nu_1 - \nu_2$ plane).  
The Hamiltonian of the system in  the  $\nu_f^{(3)}$ basis   
can be written as 
\be
H_f = U^{\dagger}_{\theta} U^{\dagger}_{\alpha} H^{diag} U_{\alpha} U_{\theta} + V,  
\ee
where 
\bea
H^{diag} & \equiv &  {\rm diag}(H_0,~H_1,~H_2) = 
\frac{1}{2E} {\rm diag}(m_0^2,~m_1^2,~ m_2^2),
\nonumber\\
V & \equiv & {\rm diag}(0, V_e, V_a) 
\eea
are the diagonal matrices of the  eigenvalues of the Hamiltonian in vacuum and 
the matter potentials correspondingly; $V_e = \sqrt{2}G_F (n_e - 0.5 n_n)$ and 
$V_a = - (1/\sqrt{2})G_F n_n$, with $n_e$ and $n_n$ being the 
electron and neutron number densities.  

It is convenient to consider effects of sterile neutrino mixing in the basis 
rotated by the 1-2 mixing in matter $U_{\theta_m}$: $(\nu_s,  \nu_{1m}^{LMA}, 
\nu_{2m}^{LMA})$, 
which would diagonalize the Hamiltonian in the absence of mixing with sterile neutrinos 
({\it i.e.} when $U_{\alpha} = I$). In this basis 
the Hamiltonian becomes 
\be
H_\alpha = U_{\theta_m} H_f U^{\dagger}_{\theta_m} = 
U_{\theta_m} U^{\dagger}_{\theta} U^{\dagger}_{\alpha} H^{diag}  U_{\alpha} U_{\theta}
U^{\dagger}_{\theta_m}  +  U_{\theta_m}V U^{\dagger}_{\theta_m}. 
\label{eq:hamalp}
\ee
Since $U_{\alpha}$ is small rotation we can represent it as 
\be
U_{\alpha} = I + U_\delta, 
\label{eq:ualpha}
\ee
where $U_\delta \equiv U_{\alpha} - I$. Inserting this expression into  
(\ref{eq:hamalp}) and taking the lowest order terms in $U_\delta$ we obtain 
\be
H_\alpha = H_m^{diag} + H_\delta, 
\label{eq:halpha}
\ee
where 
\be
H_m^{diag} = {\rm diag}(H_0, \lambda_1^{LMA}, \lambda_2^{LMA})
\ee
is the Hamiltonian in the absence of mixing with sterile neutrino and 
\be
H_\delta \equiv U^{\dagger}_{\Delta \theta} U^{\dagger}_\delta H^{diag} U_{\Delta \theta} + 
h.c. 
\label{eq:hdelt}
\ee
is the correction to the Hamiltonian due to  mixing with sterile neutrino. 
Here $\Delta \theta \equiv (\theta - \theta_m)$ and  
\be 
U_{\Delta \theta} \equiv U_{\theta} U^{\dagger}_{\theta_m} = 
\left(\begin{array}{ccc}
1 & 0 & 0\\
0 & \cos(\theta - \theta_m)  & - \sin(\theta - \theta_m)\\
0 & \sin(\theta - \theta_m) & \cos(\theta - \theta_m) 
\end{array}
\right).
\label{eq:12diff}
\ee

We denote the ratio of mass squared differences as 
$$
R_{\Delta} \equiv \frac{\Delta m^2_{01}}{\Delta m^2_{21}}.
$$ 
Depending on mass and mixing of the sterile neutrino (i.e. the form of $U_\alpha$)  
one can obtain several phenomenologically different possibilities.

\subsection{The case $m_1 < m_0 < m_2$ }

This case corresponds to  $R_\Delta \ll 1$ (as in \cite{pedroS}).  
For the neutrino with  energy $E = 10$ MeV 
the level crossing scheme which gives  dependence of
$\lambda_i$, ($i = 0, 1, 2$) on the distance inside the Sun (or
on the density), is shown in fig.~\ref{fig:eigenvalues}. 
Details of construction of this scheme can be found in \cite{pedroS}.  
For other energies the scheme can be obtained from the one in 
Fig.~\ref{fig:eigenvalues} by shifting the picture with respect to the frame
to the right with increase of energy and 
to the left with decrease of energy. 
According to fig.~\ref{fig:eigenvalues} for $E = 10$ MeV the sterile neutrino 
level, $\lambda_s$, has two resonances -- two crossings with the original 
(without $\nu_s$) level $\lambda_{1}^{LMA}$:  at smaller density $n_l^{R}$, 
and at higher density, $n_h^{R}$, ($n_h^{R} > n_l^{R}$). 
With increase of $\Delta m^2_{01}$ the density $n_l^{R}$ increases, whereas 
$n_h^{R}$ - decreases, they approach each other and then merge.

\begin{figure}[ht]
\begin{center}
\vskip 1cm
\includegraphics[width=13cm]{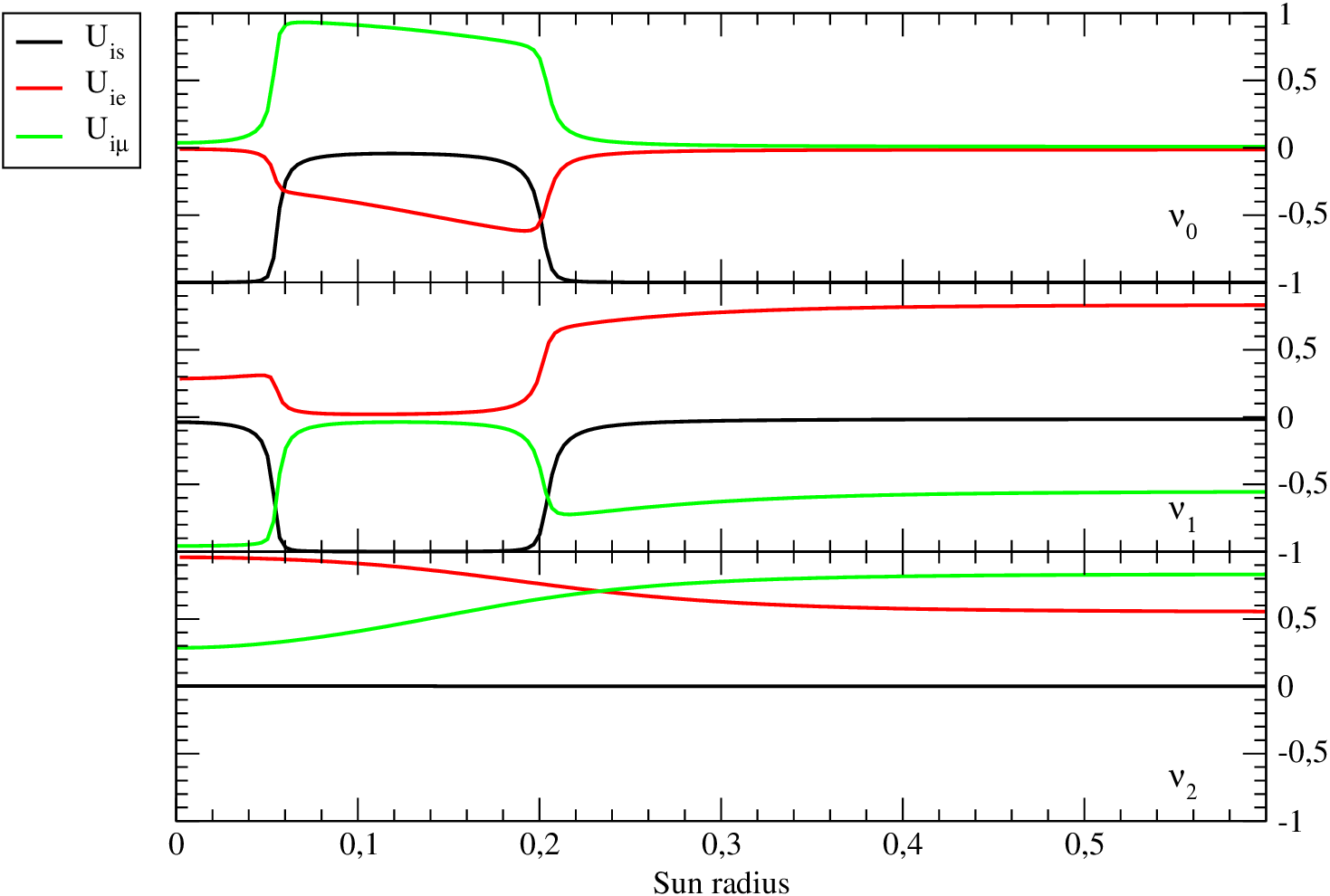}
\caption{Flavor content of the mass eigenstates in matter  
for the same neutrino parameters as in fig.~\ref{fig:eigenvalues}. }
\label{fig:mixing}
\end{center}
\end{figure}

The flavor content of the eigenstates in matter, $\nu_{im}$,   
{\it i.e.},  the mixing matrix elements $U_{i\alpha}^m(n(r))$ as functions 
of density (distance from the center of the Sun)  
is shown in fig~\ref{fig:mixing}.  Notice that since the  
$\nu_s$ mixing in the $\nu_2$ is absent, 
the change of flavor of $\nu_{2m}$  with density  
is the same as in the LMA case:  $U_{2\alpha}^m = U_{2\alpha}^{mLMA}$. 
The intersections of the lines correspond to resonances.    

Consider evolution of this system from the production
point in the central region of the Sun to the solar surface. The eigenstates 
$\nu_{2m}$ and $\nu_{3m}$ evolve adiabatically  
$$
\nu_{2m} \rightarrow \nu_{2}, ~~~ \nu_{3m} \rightarrow \nu_{3},   
$$
{\it i.e.}, $A_{2j} = \delta_{2j}$,   $A_{3j} = \delta_{3j}$ and 
therefore according to (\ref{eq:amplij})
\be 
A_{e2}^S = U_{e2}^m,  ~~~ A_{e3}^S = U_{e3}^m \approx U_{e3}.  
\label{eq:amp23}
\ee
In the last equality we have taken into account that   
due to large value of $\Delta m_{31}^2$ 
the 1-3 mixing is practically unaffected by the solar matter 
in the energy range of interest.
Inserting the amplitudes (\ref{eq:amp23}) 
into (\ref{eq:gen1}) we obtain 
\be
P_{ee} = |A_{e1}^S|^2 |U_{e1}|^2 + |A_{e0}^S|^2 |U_{e0}|^2 + 
|U_{e2}^m|^2 |U_{e2}|^2 + |U_{e3}|^4. 
\label{eq:gen2}
\ee
In turn, the amplitudes of $\nu_e-$transitions 
to $\nu_1$ and to $\nu_0$ can be written 
according to (\ref{eq:amplij}) as 
\be
A_{e1}^S  =  U_{e1}^m A_{11} + U_{e0}^m A_{01}, ~~~~~
A_{e0}^S  =  U_{e1}^m A_{10} + U_{e0}^m A_{00}.  
\label{eq:amplij1}
\ee
Finally, insertion of (\ref{eq:amplij1}) into (\ref{eq:gen2}) gives 
\be
P_{ee} = |U_{e1}^m A_{11} + U_{e0}^m A_{01}|^2 |U_{e1}|^2 + 
|U_{e1}^m A_{10} + U_{e0}^m A_{00}|^2 |U_{e0}|^2 +
|U_{e2}^m|^2 |U_{e2}|^2 + |U_{e3}|^4.
\label{eq:pee1}
\ee
Since $|U_{e0}|^2 \sim 10^{-3}$ and $|U_{e3}|^4 \sim 4 \cdot 10^{-4}$, 
the probability (\ref{eq:pee1}) can be written approximately as 
\be
P_{ee} \approx |U_{e1}^m A_{11} + U_{e0}^m A_{01}|^2 |U_{e1}|^2 +
|U_{e2}^m|^2 |U_{e2}|^2 .
\label{eq:pee2}
\ee

With decrease of neutrino energy effectively
the pattern in fig. ~\ref{fig:eigenvalues} shifts to the left. 
Therefore at low energies there is no sterile neutrino resonances, 
the evolution proceeds adiabatically and 
\be
P_{ee} \approx |U_{e1}^m|^2 |U_{e1}|^2 +
|U_{e2}^m|^2 |U_{e2}|^2 .
\label{eq:pee3}
\ee
Furthermore,  $U_{e1}^m \approx U_{e1}^{mLMA}$ and $U_{e2}^m \approx U_{e2}^{mLMA}$,  
where  $U_{ei}^{mLMA}$  are mixing parameters in the pure LMA 
($2\nu$) case without sterile neutrinos.  So, $P_{ee} \approx P_{ee}^{LMA}$. 

With increase of energy the low density resonance becomes effective 
(the corresponding level crossing scheme for $E = 8$ MeV 
is shown in fig.~1 of our paper \cite{pedroS}).  
In the adiabatic case (relatively large $U_{01}$) we have  $A_{01} \approx 0$, $A_{11} \approx 1$ 
and the expression for $P_{ee}$ in eq.~(\ref{eq:pee2}) 
is reduced to the one  in (\ref{eq:pee3}). 
However  now $U_{e1}^m \approx 0$  if neutrino is produced above 
the resonance layer, and consequently, $P_{ee} \approx |U_{e2}^m|^2 |U_{e2}|^2$. 
 
In the non-adiabatic case: 
\be
P_{ee} \approx |U_{e0}^m A_{01}|^2 |U_{e1}|^2 +
|U_{e2}^m|^2 |U_{e2}|^2 \approx |U_{e1}^{mLMA} A_{01}|^2 |U_{e1}|^2 +
|U_{e2}^m|^2 |U_{e2}|^2,  
\label{eq:pee4}
\ee
where we have taken into account that  $U_{e0}^m \approx  U_{e1}^{mLMA}$.
In the case of strong adiabaticity violation (for very small $U_{01}$) 
when $A_{01} \approx 1$, the probability in (\ref{eq:pee4}) is reduced to the 
standard LMA probability.   

With further increase of energy at $E \aprge 9$ MeV 
two sterile resonances are realized and the amplitudes $A_{01}$ and $A_{11}$ 
can be written in terms of transition amplitudes in each resonance 
$A_{ij}^{(a)}$ ($a = 1,2$) as
\be 
A_{11}  =  A_{11}^{(2)} A_{11}^{(1)} +  A_{10}^{(2)} A_{01}^{(1)}, ~~~~~
A_{01}  =  A_{00}^{(2)} A_{01}^{(1)} +  A_{01}^{(2)} A_{11}^{(1)}.  
\label{eq:relation}
\ee  
One can get different results depending on the adiabaticity 
conditions in each resonance. 
If the crossings are adiabatic, 
$A_{01} \approx 0$, $A_{11} \approx 1$, we obtain
from (\ref{eq:pee2})
$$
P_{ee} \approx |U_{e1}^m|^2 |U_{e1}|^2 + |U_{e2}^m|^2 |U_{e2}|^2 .
$$
This expression is similar to the one for the standard LMA case, however  now 
$
|U_{e1}^m|^2 = 1 - |U_{e2}^m|^2 - |U_{e0}^m|^2 = 
|U_{e1}^{mLMA}|^2 -|U_{e0}^m|^2,  
$
and consequently, 
$$
P_{ee} = P_{ee}^{LMA} - |U_{e1}|^2 |U_{e0}^m|^2. 
$$ 
Here the second term describes the dip in the adiabatic case. 
(Notice that due to smallness of $\nu_s$ mixing in vacuum 
$U_{e1} \approx U_{e1}^{LMA}$.)

If adiabaticity is strongly broken in both resonances due to smallness 
of sterile mixing, then  $A_{10}^{(2)} \approx A_{01}^{(1)} \approx 1$ 
and  $A_{11}^{(2)} \approx A_{00}^{(2)} \approx A_{11}^{(1)} \approx 0$.   
Therefore according to (\ref{eq:relation}),  
$A_{01} \approx 0$, $A_{11} \approx 1$, and as in the adiabatic case:   
\be
P_{ee} \approx |U_{e1}^m|^2 |U_{e1}|^2 + |U_{e2}^m|^2 |U_{e2}|^2 .
\label{eq:nonadiab}
\ee
However, now the mixing parameters are approximately equal to 
the standard LMA parameters without sterile neutrinos. 
Therefore $P_{ee} \approx P_{ee}^{LMA}$. 

In the case of mixing of $\nu_s$ in $\nu_1$ and  $\nu_0$ only, the matrix 
$U_\alpha$ equals
\be
U_\alpha = 
\left(\begin{array}{ccc}
\cos \alpha &  \sin \alpha  &  0\\
- \sin \alpha   & \cos \alpha  & 0\\
0 & 0  & 1
\end{array}
\right).
\label{eq:alpha01}
\ee
Explicitly the flavor  mixing can be parameterized as 
\begin{eqnarray}
\nu_0 &=& \cos \alpha~\nu_s + \sin \alpha (\cos\theta_{12}~\nu_e  - \sin\theta_{12}~\nu_a), 
\nonumber\\
\nu_1 &=&  \cos \alpha~ (\cos\theta_{12}~\nu_e  - \sin\theta_{12}~\nu_a) - \sin \alpha~\nu_s, 
\nonumber\\
\nu_2 &=& \sin\theta_{12}~\nu_e + \cos\theta_{12}~\nu_a~.
\nonumber
\end{eqnarray}
Here $\nu_e$ and $\nu_a$ (a combination of $\nu_\mu$ and $\nu_\tau$) 
mix with the angle $\theta_{12}$ 
in the mass eigenstates $\nu_1$ and $\nu_2$ having 
the mass split $\Delta m^2_{12}$. 
In terms of these mixing angles: 
$$ 
U_{e1} = \cos \alpha~\cos\theta_{12}, ~~~ 
U_{e0} = \sin \alpha~\cos\theta_{12}, ~~~ U_{e2} = \sin \theta_{12}. 
$$
With this parametrization eqs. (\ref{eq:pee2}) and (\ref{eq:nonadiab}) 
reproduce the corresponding results of our previous paper \cite{pedroS}. 

{} From eqs. (\ref{eq:halpha} - \ref{eq:12diff}) we find using 
(\ref{eq:alpha01}) that the Hamiltonian in the rotated basis 
equals
\be
H_{\alpha} = 
\left(\begin{array}{ccc}
0 &  -\frac{\Delta m_{01}^2}{2 E} \sin \alpha  \cos (\theta - \theta_m) &                       
  -\frac{\Delta m_{01}^2}{2 E} \sin \alpha  \sin (\theta - \theta_m)\\
...   & \lambda_1^{LMA} - H_0  & 0\\
... & ...  & \lambda_2^{LMA} - H_0 
\end{array}
\right).
\label{eq:halpha1}
\ee
Due to smallness of $\alpha$ the off-diagonal terms are much smaller than the diagonal ones. 
If $R_\Delta \ll 1$, so that $H_0$ is close to $\lambda_1^{LMA}$ and crosses this level,   
there is no  
crossing of $\lambda_2^{LMA}$ and  $H_0$ levels, and 
the state $\nu_{2m}^{LMA}$ decouples. 
Then mixing of $\nu_s$ and $\nu_{1m}^{LMA}$ is determined by 
\be
\sin \alpha  \frac{\Delta m_{01}^2}{2 E}  \cos (\theta - \theta_m)
=  \sin \alpha \frac{\Delta m_{21}^2}{2 E} R_\Delta \cos(\theta - \theta_m).  
\label{eq:mixing-01}
\ee
Since the transition occurs in the resonance region,   
$\theta_m$ should be taken at the density which corresponds to the sterile 
neutrino resonance. 
The expression for mixing (\ref{eq:mixing-01}) allows to understand behavior of 
the conversion probability on $R_\Delta$,  $\alpha$ and neutrino energy.

\begin{figure}[ht]
\begin{center}
\vskip 1.5cm
\includegraphics[width=13cm]{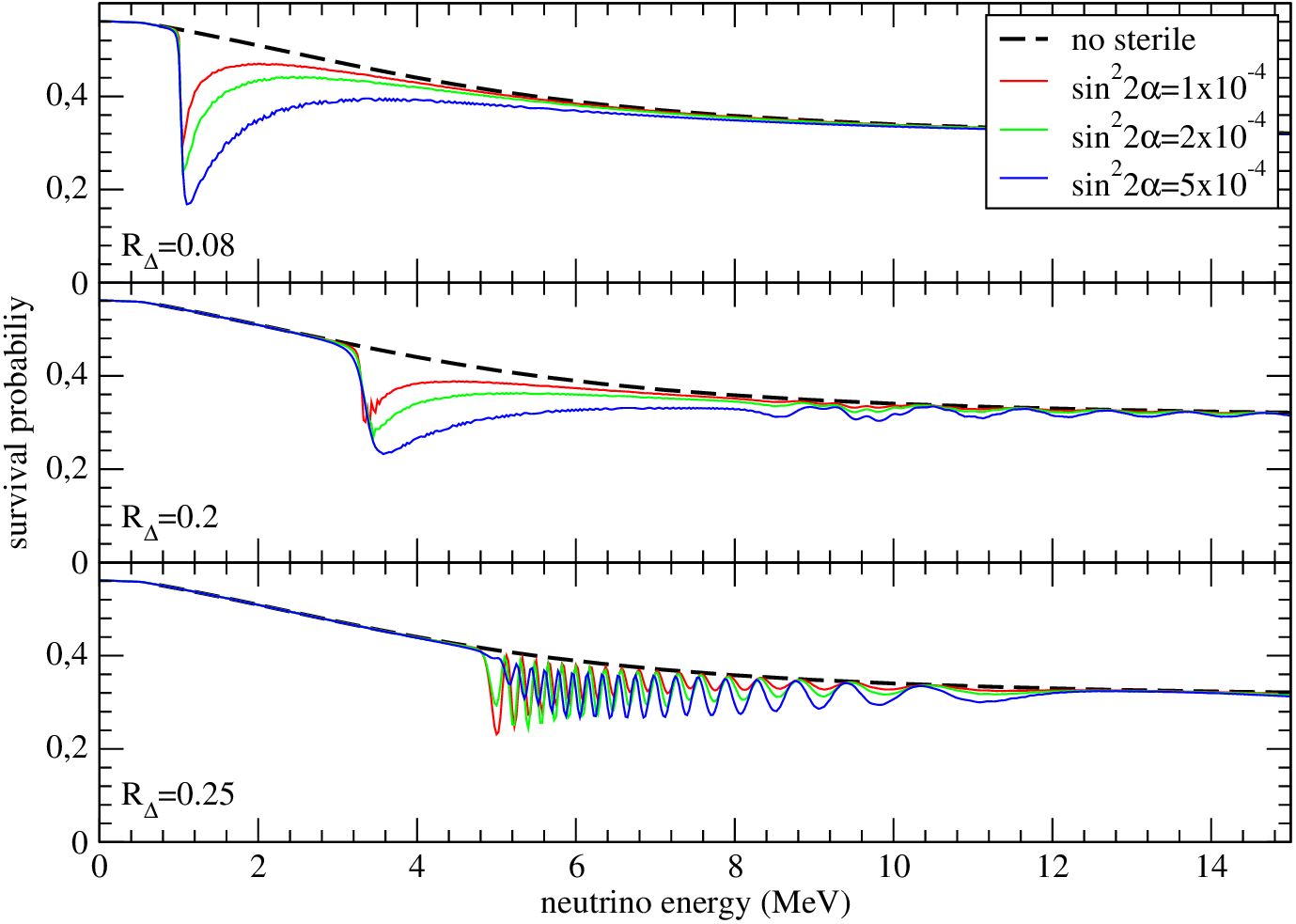}
\caption{
The survival probability of the electron neutrino 
as function of neutrino energy
for different values of the sterile-active mixing  
parameter $\sin^2 2\alpha$ and mass scale $R_\Delta \ll 1$. 
The active neutrino parameters are $\Delta m^2_{21}=8\times
10^{-5}$ eV$^2$ and $\tan^2\theta=0.44$.}
\label{fig:psurv}
\end{center}
\end{figure}

\begin{figure}[ht]
\begin{center}
\vskip 1.5cm
\includegraphics[width=13cm]{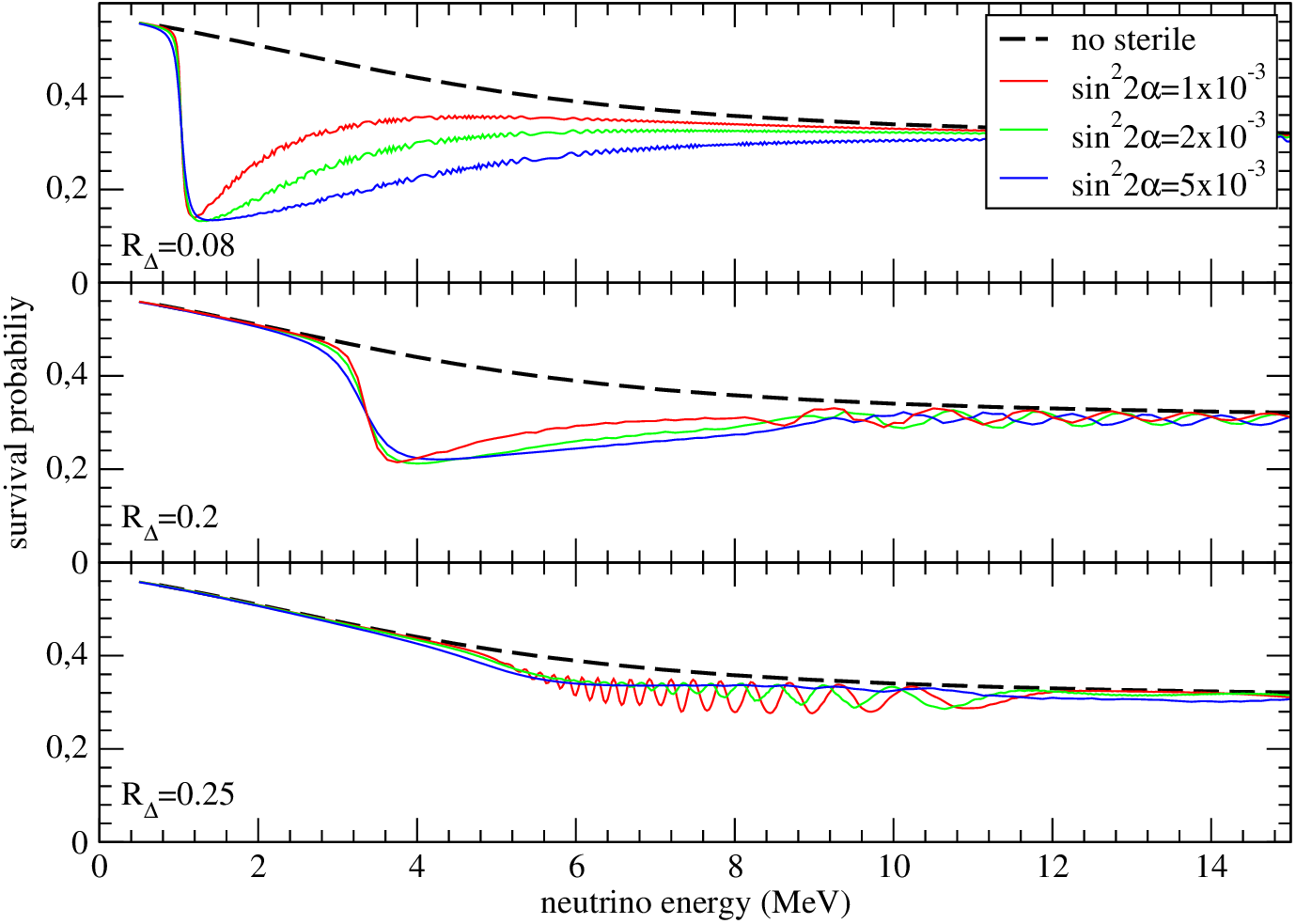}
\caption{The same as in fig.~\ref{fig:psurv} 
for higher values of mixing angle $\alpha$.}
\label{fig:psurvh}
\end{center}
\end{figure}

In figs.~\ref{fig:psurv},  \ref{fig:psurvh}   we show results of numerical
computations of the $\nu_e$ survival probability $P_{ee}$ as
functions of the neutrino energy.  We have performed
complete integration of the evolution equations for the $3\nu$-system
and also made  averaging over the production region of
the Sun.  The following analytical consideration allows to understand
the numerical results shown in fig. \ref{fig:psurv}.
The $\nu_s-$mixing and $\nu_e \rightarrow \nu_s$ conversion  lead to appearance of a 
dip in the energy dependence of the $\nu_e-$ survival probability. 
The survival probability in the pure LMA case is given by 
\be
P_{ee}^{LMA} \approx |U_{e1}^{mLMA}|^2 |U_{e1}^{LMA}|^2 + |U_{e2}^{mLMA}|^2 |U_{e2}^{LMA}|^2 + 
|U_{e3}|^4. 
\label{eq:eelma}
\ee
Since $\nu_s$ mixing in the $\nu_2$ is absent, 
$U_{e2}^{LMA} = U_{e2}$ and 
$U_{e2}^{mLMA} = U_{e2}^m$ and the probability (\ref{eq:eelma})
can be rewritten as 
\be
P_{ee}^{LMA} \approx |U_{e1}^{0m}|^2 |U_{e1}^0|^2 + |U_{e2}^{m}|^2 |U_{e2}|^2 + 
|U_{e3}|^4.
\label{eq:eelma1}
\ee
Using normalization conditions
$$
|A_{e1}^S|^2 + |A_{e0}^S|^2 +  |U_{e2}^m|^2  = 1 
$$
(we neglect 1-3 mixing here)
and $\sum_i |U_{ei}|^2 = 1$ (i = 0,1,2,3) as well as 
the expressions in eqs. (\ref{eq:gen2}) and (\ref{eq:eelma1})
we can find difference of the probabilities 
with and without sterile neutrino effect: 
\begin{eqnarray}
\Delta P_{ee} & \equiv & P_{ee}^{LMA} - P_{ee}  = 
|A_{e0}^S|^2 (1 - |U_{e2}|^2) - 
|U_{e0}|^2 (2|A_{e0}^S|^2 - 1 - |U_{e2}^m|^2)
\nonumber\\ 
& \approx & |A_{e0}^S|^2 (1 - |U_{e2}|^2) \approx P_{es} (1 - |U_{e2}|^2), 
\nonumber
\end{eqnarray}
where $P_{es} \approx |A_{e0}^S|^2$ is the probability of 
$\nu_e \rightarrow \nu_s$ transition. 
The quantity $\Delta P_{ee}$ describes the dip which  
has the following properties (see fig.~\ref{fig:psurv} and also discussion in \cite{pedroS}):  

1. A position of the dip (its low energy edge) is given by the low density resonance 
taken at the central density of the Sun $E_l (n_c)$. 
With increase of $\Delta m_{01}^2$ the dip shifts
to higher energies. 

2. Maximal suppression in the dip depends on $R_{\Delta}$ and 
$\alpha$. For small $R_{\Delta}$ (large spit between the two
resonances) and large $\alpha$ ($\sin^2 2\alpha > 10^{-3}$) the
absolute minimum can be achieved at the adiabatic crossing 
of the $\nu_s-$resonances. 
With increase of $R_{\Delta}$ (smaller split of the resonances) or/and
decrease of  $\alpha$ (stronger violation of the adiabaticity) a 
suppression in the dip weakens.  Also with decrease of $\alpha$ 
the dip becomes narrower.

3. For large $\Delta m^2_{01}$  and relatively small $\alpha$
the survival probability as function of the neutrino energy 
has  wiggles (see fig. \ref{fig:psurv}). The wiggles are result of 
interference 
of the two amplitudes  in  the first term of (\ref{eq:pee2}) which develops over finite 
space interval.    
Indeed, according to (\ref{eq:pee2}) there are two channels of transition 
of $\nu_e$ to $\nu_1$: 

(i) $\nu_e$ has admixture  $U_{e1}^m$ in $\nu_{1m}$, the latter adiabatically evolves to 
$\nu_{1}$: $\nu_e \rightarrow \nu_{1m} \rightarrow \nu_{1}$, and the amplitude equals   
$U_{e1}^m A_{11}$.  

(ii) $\nu_e$ has admixture $U_{e0}^m$ in $\nu_{0m}$; this state  
transforms to $\nu_{1}$ due to non-adiabatic transition: 
$\nu_e \rightarrow \nu_{0m} \rightarrow \nu_{1}$. The corresponding amplitude 
is $U_{e0}^m A_{01}$.   

The two contributions to the amplitude interfere leading to the
oscillatory dependence of the probability on energy (wiggles). 
Introducing $P_{01} \equiv |A_{01}|^2$, so that $|A_{11}|^2 = 1 - P_{01}$,   
we can rewrite the probability (\ref{eq:pee2}) as 
$$
P_{ee} \approx |U_{e1}|^2 \left[|U_{e1}^m|^2 (1 - P_{01}) + |U_{e0}^m|^2 P_{01}
+ U_{e1}^m U_{e0}^m \cos \phi \sqrt{P_{01} (1 - P_{01})}\right] + 
|U_{e2}|^2 |U_{e2}^m|^2, 
$$
where $\phi \equiv arg(A_{01}^*A_{11})$ and we assumed for 
simplicity that  $U_{e1}^m$ and  $U_{e0}^m$ are real. The oscillatory behavior 
follows from the energy dependence of the phase $\phi$. The key point is that 
the phase is collected over restricted space interval, $L$, and therefore 
is not averaged out even after integration over the production region. 
Indeed, the phase $\phi$ is acquired from the neutrino production point to 
the second (low density) resonance. 
Below the second resonance (in density) both ``trajectories'' 
(channels of transition) coincide. Appearance of the wiggles requires 
the adiabaticity violation. 
In the adiabatic case $A_{01} = 0$ and only one channel exists. 
Unfortunately, it will be difficult, if possible, to observe experimentally 
these wiggles. Some more details concerning the wiggles are presented 
in the Appendix. 

If $\nu_s$ mixes in $\nu_2$, then
\be
U_\alpha = 
\left(\begin{array}{ccc}
\cos \alpha^\prime &  0 & \sin \alpha^\prime \\
0 & 1  & 0 \\
- \sin \alpha^\prime   & 0 & \cos \alpha^\prime 
\end{array}
\right)
\label{eq:alpha02}
\ee
and the Hamiltonian can be obtained from (\ref{eq:halpha1}) by substitutions  
\be 
\cos (\theta - \theta_m) \rightarrow \sin (\theta - \theta_m), ~~~
\sin (\theta - \theta_m) \rightarrow - \cos (\theta - \theta_m), ~~~
\Delta m^2_{01} \rightarrow \Delta m^2_{02}. 
\label{eq:subst}
\ee 
Again the state $\nu_{2m}$ decouples and $\nu_s - \nu_{1m}^{LMA}$ mixing is given by 
\be
 \sin \alpha^\prime \frac{\Delta m_{02}^2}{2 E} \sin(\theta - \theta_m)
=  \sin \alpha^\prime \frac{\Delta m_{21}^2}{2 E} (1 - R_\Delta) 
\sin (\theta - \theta_m).
\label{eq:mixing-01pr}
\ee
Notice that this mixing appears due to matter effect and it is absent in vacuum 
when $\theta_m \rightarrow \theta$. It happens that for values of 
$R_\Delta$ we are considering $(1 - R_\Delta) \sin (\theta - \theta_m) \approx 
R_\Delta \cos (\theta - \theta_m)$ and therefore the 
probabilities in this case  are very similar to those shown in figs.  
~\ref{fig:psurv},  \ref{fig:psurvh}.

\subsection{$m_0 > m_2 > m_1$ and other possibilities}

For  $m_0 > m_2 > m_1$ the ratio  $R_{\Delta} > 1$. Since below the LMA resonance 
$\lambda_2 = \lambda_2^{LMA}$ and  $V_n$ have practically the same 
dependences on density (radius) (see fig.~\ref{fig:eigenvalues}),
there is only one crossing 
of $\lambda_s$ with $\lambda_2^{LMA}$, and there is no crossings 
for $\Delta m_{02}^2 < 0$. 
Now the evolution of states $\nu_{1m}$ and $\nu_{3m}$ is adiabatic, so that 
$$
A_{e1}^S \approx U_{e1}^m = U_{e1}^{0m}, ~~~~ A_{e3}^S \approx U_{e3}.  
$$  
Consequently, 
$$
P_{ee} = |U_{e1}^m|^2 |U_{e1}|^2 + |A_{e2}^S|^2 |U_{e2}|^2 + 
|A_{e0}^S|^2 |U_{e0}|^2  + |U_{e3}|^4,  
$$
where 
$$
A_{e2}^S  =  U_{e2}^m A_{22} + U_{e0}^m A_{02}, ~~~~~
A_{e0}^S  =  U_{e2}^m A_{20} + U_{e0}^m A_{00}.  
$$
These expressions are similar to the expressions in (\ref{eq:gen2}) and 
(\ref{eq:amplij}) with interchange of indexes $1 \leftrightarrow 2$.  

In the adiabatic case we have 
$$
P_{ee} = |U_{e1}^m|^2 |U_{e1}|^2 + |U_{e2}^m|^2 |U_{e2}|^2 + 
|U_{e0}^m|^2 |U_{e0}|^2  + |U_{e3}|^4. 
$$
Now the effect of sterile neutrino is due to difference of 
$U_{e2}^m$ and   $U_{e2}^{mLMA}$: $|U_{e2}^m|^2 = |U_{e2}^{mLMA}|^2 - |U_{e0}^m|^2$. 
In the strongly non-adiabatic case $A_{20} \approx A_{02} \approx 1$  and 
\be
P_{ee} = |U_{e1}^m|^2 |U_{e1}|^2 + |U_{e2}^m|^2 |U_{e0}|^2 + 
|U_{e0}^m|^2 |U_{e2}|^2  + |U_{e3}|^4,  
\label{eq:nonpee}
\ee
with $U_{e2}^m \approx 0$, and  $U_{e0}^m \approx U_{e2}^{mLMA}$, so that 
(\ref{eq:nonpee}) is reduced to the LMA probability. 

\begin{figure}[ht]
\begin{center}
\vskip 1.5cm
\includegraphics[width=13cm]{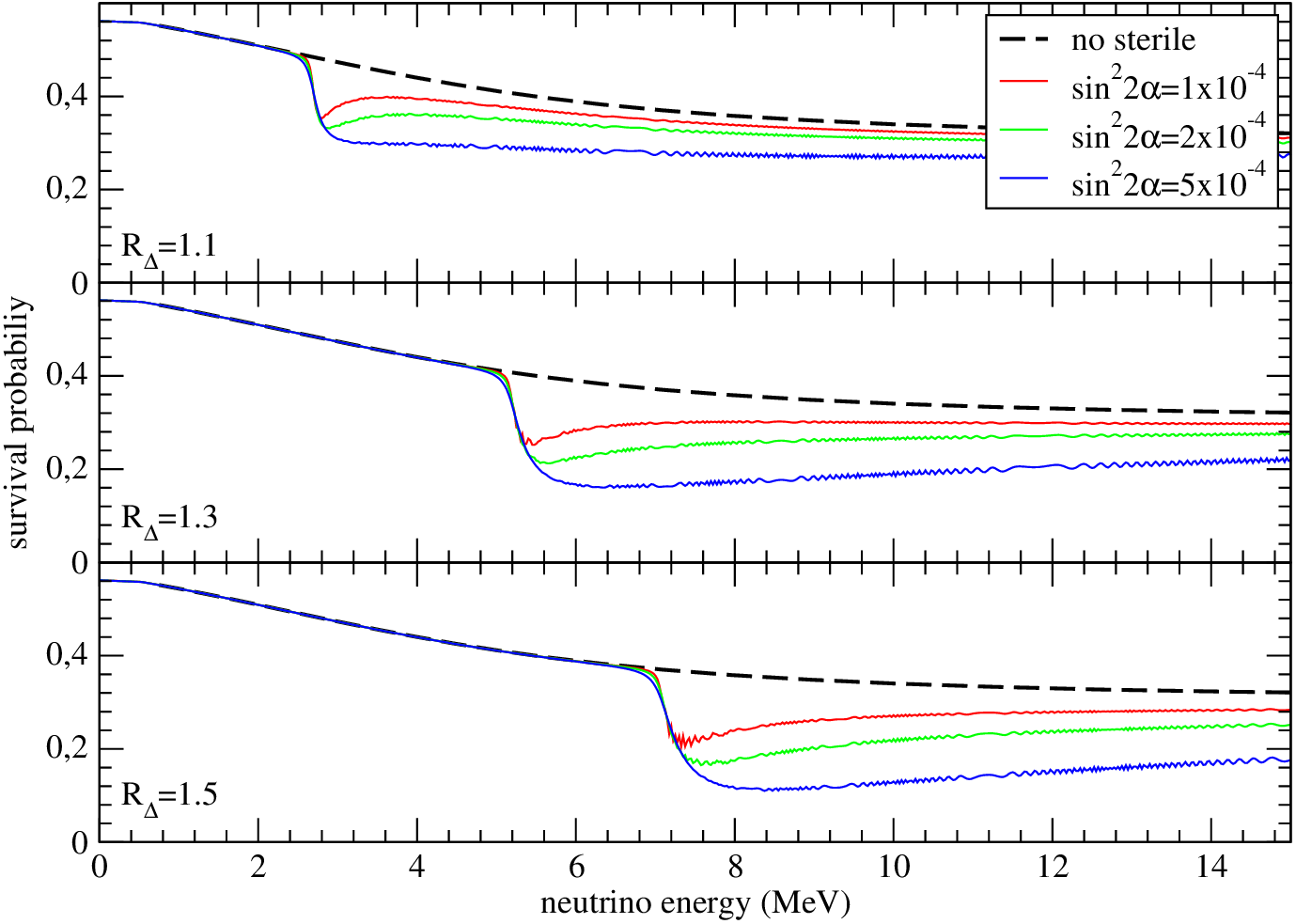}
\caption{
The same as in fig.~\ref{fig:psurv} for $R_\Delta > 1$. 
}
\label{fig:psurvb}
\end{center}
\end{figure}

\begin{figure}[ht]
\begin{center}
\vskip 1.5cm
\includegraphics[width=13cm]{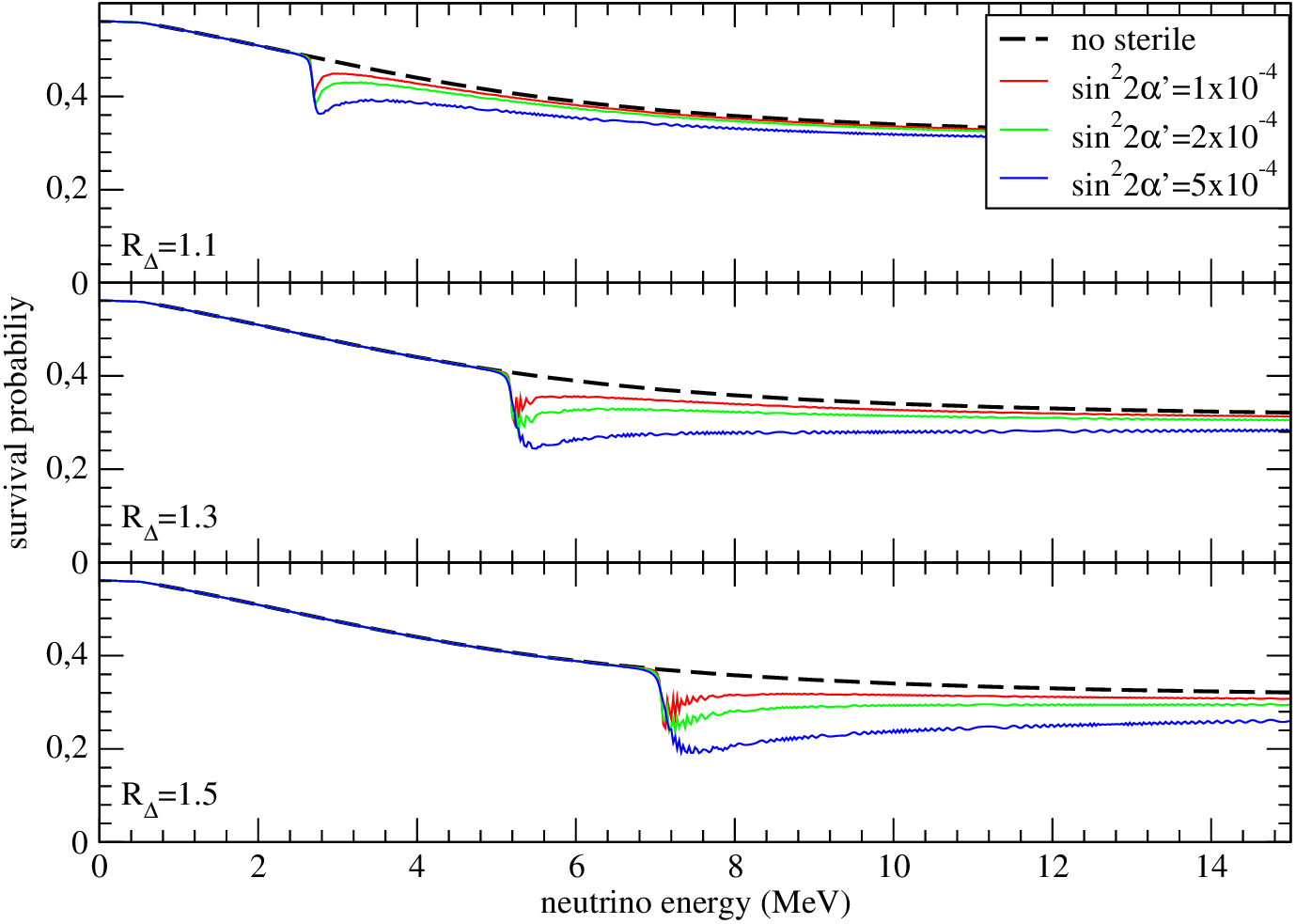}
\caption{
The same as in fig.~\ref{fig:psurvh} for $R_\Delta > 1$ 
and $\nu_s$ mixing in $\nu_2$.
}
\label{fig:psurvc}
\end{center}
\end{figure}

For $\nu_s-$mixing in $\nu_0$ and $\nu_1$ the Hamiltonian is given 
by the same expression as in eq. (\ref{eq:halpha1}). 
However, now  $H_0$ crosses $\lambda_2^{LMA}$ and 
the state $\nu_{1m}^{LMA}$ decouples. 
According to (\ref{eq:halpha1}), the mixing of $\nu_s$ and $\nu_{2m}^{LMA}$ 
is determined by
\be
\sin \alpha  \frac{\Delta m_{01}^2}{2 E}  \sin (\theta - \theta_m)
=  \sin \alpha \frac{\Delta m_{21}^2}{2 E} R_\Delta \sin(\theta - \theta_m),  
\label{eq:mixing-01l}
\ee
and this mixing is due to matter effect. 
For $\nu_s-$mixing in $\nu_0$ and $\nu_2$ performing 
substitutions (\ref{eq:subst}) we obtain the 
$\nu_s - \nu_{2m}^{LMA}$ mixing 
\be
\sin \alpha^{\prime}  \frac{\Delta m_{02}^2}{2 E}  \cos (\theta - \theta_m)
=  \sin \alpha^{\prime} \frac{\Delta m_{21}^2}{2 E} (1 - R_\Delta) \cos (\theta - \theta_m).
\label{eq:mixing-02l}
\ee
In the first case, eq.~(\ref{eq:mixing-01l}),  
the mixing  $\propto  R_\Delta \sin(\theta - \theta_m)$ 
is larger than in second case (\ref{eq:mixing-02l}): 
$\propto (1 - R_\Delta) \cos (\theta - \theta_m)$, since $R_\Delta \sim 1$. 
Furthermore, the first mixing increases with energy: 
the sterile resonance is above the LMA resonance and therefore
$\theta_m > 45^{\circ}$; this angle, and consequently $|\sin(\theta - \theta_m)|$,  
increase.  As a result, the effect does not disappear at high energies
(see \ref{fig:psurvb} and \ref{fig:psurvc}).

In comparison to the case $R_\Delta \ll 1$, now   
$\nu_e$ has smaller admixture in $\nu_2$, 
$|U_{e2}| < |U_{e1}|$,  
however the initial admixture of $\nu_e$ in $\nu_{2m}$ can be  larger: 
$|U_{e2}^m| >  |U_{e1}^m|$.  Therefore the overall effect is large (see figs~
\ref{fig:psurvb} and \ref{fig:psurvc}). 
Indeed here we have only one level crossing and improvement of the adiabaticity 
in the resonance leads to stronger transition. With increase of 
$\Delta m_{02}^2$ and therefore,  $R_\Delta$, the dip moves to high energies but 
the resonance shifts to higher densities,  {\it i.e.},  to the central regions of the Sun 
where the density gradient is smaller and adiabaticity is better. 
Here substantial  change of the probability  exists for smaller mixing angles. 

If $m_0 < m_1 < m_2$,  so that $\Delta m^2_{01} < 0$,   
the sterile level $\lambda_s$ crosses $\nu_{1m}^{LMA}$ at high densities 
only 
$$ 
n^{R}_h \approx 2 n^{LMA} (1 + R_\Delta). 
$$
The resonance energy equals 
$$
E \approx E_R^{LMA} \frac{2 n_e}{n_n} , 
$$
where $n_n$ is the number density of neutrons. 
In this case we have the same general expressions for 
the survival probability as in (\ref{eq:pee1}) and (\ref{eq:pee2}). 
Consequently, the expressions for $P_{ee}$ in adiabatic and non-adiabatic limits 
coincide with those in e.g. (\ref{eq:pee4}) for one sterile resonance. 
However the dip here is at high energies.\\

In the case of flavor mixing, that is the mixing  
of $\nu_s$ with $\nu_e$, $\nu_a$ the matrices 
$U_{\theta}$ and  $U_{\alpha}$ should be permuted,  so that 
$U^{(3)} = U_{\alpha} U_{\theta}$  
(compare with (\ref{eq:3mix})). It can be shown that now 
the off-diagonal  elements of the Hamiltonian $H_\alpha$ contain terms 
with $\Delta m^2_{01}$ and $\Delta m^2_{02}$ simultaneously. As a result, 
the probabilities have energy dependences which are intermediate between
those for mixings in  mass states $\nu_1$ and $\nu_2$.  


\section{Solar neutrino data and sterile neutrino effect}

In what follows we will consider scenario with 
$m_1 < m_0 < m_2$. This possibility  
gives better description of the data: it leads to  significant 
modification of the survival probability 
in the transition region and weakly affects spectra at high energies.  

\subsection{Borexino measurements of the Be-neutrino line}  

\begin{figure}[ht]
\begin{center}
\vskip 1cm
\includegraphics[width=12cm]{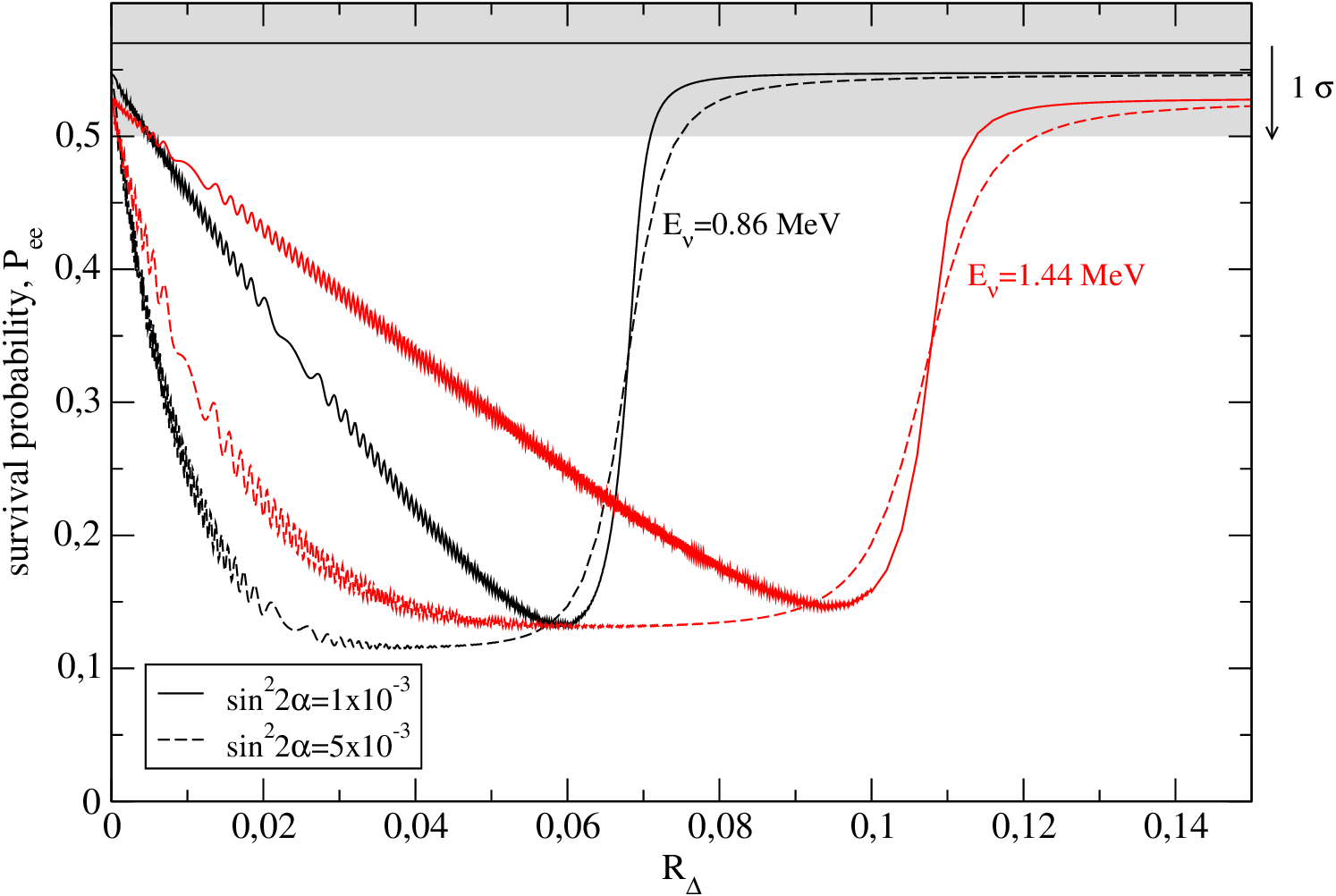}
\caption{
The survival probabilities for the monoenergetic $^7Be-$(red) 
and $pep-$ (black) neutrino fluxes  
as functions of $R_\Delta$, for 
two values of mixing:  $\sin^2 2\alpha=1\times 10^{-3}$ (solid lines) and 
$5\times 10^{-3}$ (dashed lines).  
The active neutrino oscillation parameters are the same as in 
fig.~\ref{fig:psurv}.  The horizontal line and shadowed band show 
the central value and 
$1\sigma$ band for  the suppression factor 
determined by Borexino.}
\label{fig:pmonoenergetic}
\end{center}
\end{figure}

The results of Borexino experiment \cite{Arpesella:2008mt} are in very good agreement 
with prediction based on the LMA solution and the Standard Solar Model. Within the error bars 
no additional suppression of the flux has been found on the top of $P_{ee}^{LMA}$. 
In Borexino (and other experiments based on the $\nu e$-scattering)  
the ratio of the numbers of
events  with and without conversion can be written as
\be
R_{Borexino} = P_{ee} (1 - r) + r - r P_{es},
\label{bore}
\ee
where 
$r \equiv \sigma(\nu_{\mu} e - \nu_{\mu} e)/ 
\sigma(\nu_{e} e - \nu_{e} e)$ is the ratio
of cross-sections.  Using
Eq. (\ref{bore}) we find an additional suppression 
of the Borexino rate in comparison with the pure LMA case \cite{pedroS}:
$$
\Delta R_{Borexino} \equiv  R_{Borexino}^{LMA} - R_{Borexino}  =  (1 - r)\Delta P_{ee}  + r P_{es} 
\approx  \Delta P_{ee} (1 + r \tan^2 \theta_{12}) .  
$$
In fig.~\ref{fig:pmonoenergetic} we show dependence of the 
survival probability at $E = E_{Be}$ as function 
of $R_\Delta$ for two different values of the mixing
angle $\alpha$. We show  the Borexino bounds on 
$P_{ee}$ obtained from the 
experimental result \cite{Bellini:2008mr} and relation (\ref{bore}). 
According to this figure for $\sin^2 2\alpha = 10^{-3}$ the range 
$R_\Delta  = 0.005 - 0.072$ is excluded at $1\sigma$ level.  
For $\sin^2 2\alpha = 5\cdot 10^{-3}$ we obtain slightly larger 
exclusion interval:  $R_\Delta  = 0.001 - 0.075$. 
The $Be-$neutrino line can not be in the dip or the dip should be 
shallow which then will have little impact on the higher energy spectrum.  
So, essentially the allowed values of masses  
(which influence the upturn) are 
$$
R_\Delta  \geq 0.075,~~ {\rm or} ~~  \Delta m_{01}^2 \geq 0.5 \cdot 10^{-5} {\rm eV}^2 .   
$$

\subsection{Upturn of the boron neutrino spectrum} 

\begin{figure}[ht]
\vskip 2cm 
\begin{center}
\includegraphics[width=13cm]{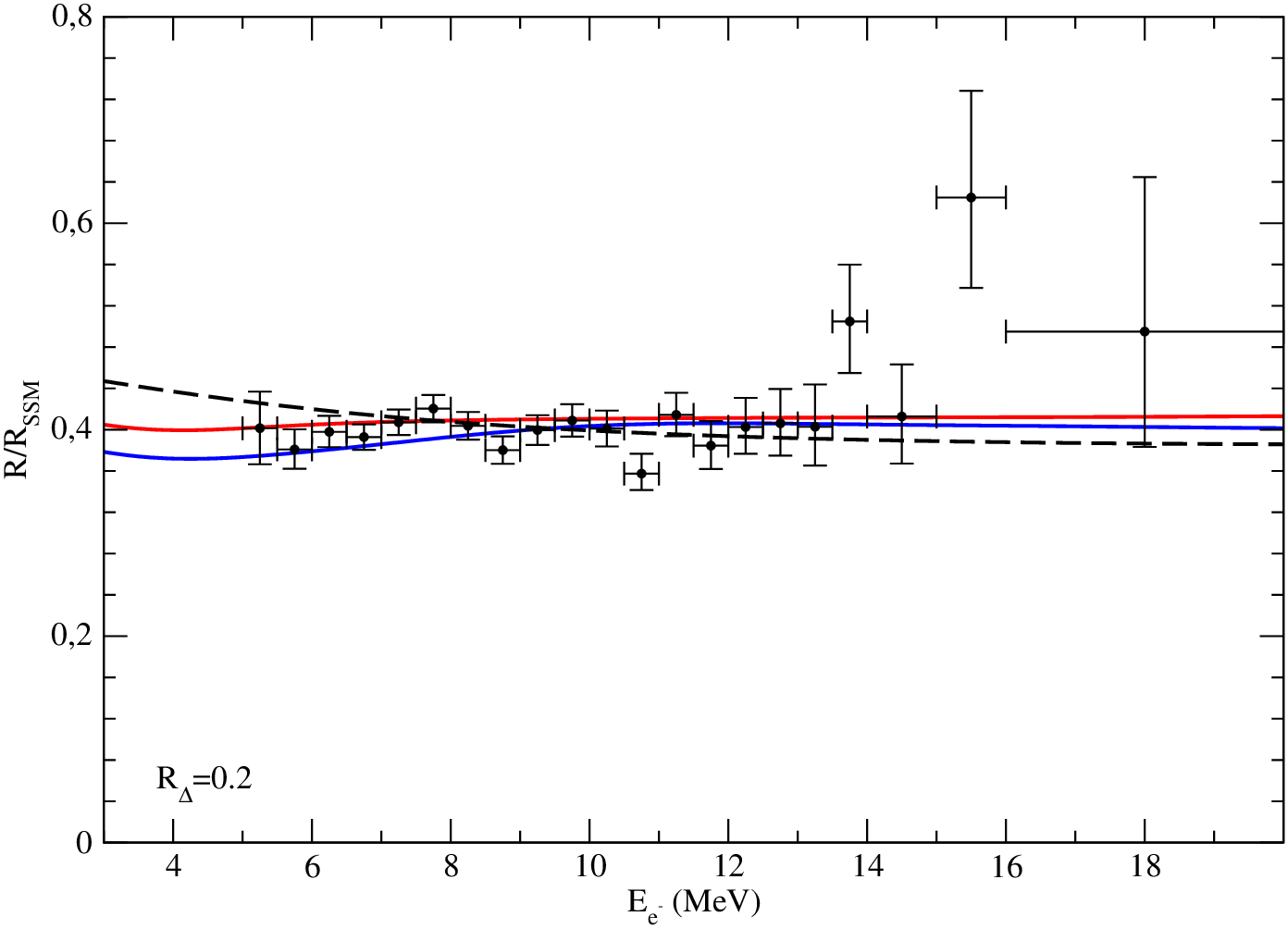}
\caption{Prediction for the SuperKamiokande-I spectrum versus  
experimental data \cite{Hosaka:2005um}. The active neutrino parameters 
are the same as before; the sterile neutrino parameters 
equal $R_\Delta=0.20$ 
and $\sin^22\alpha=1\times 10^{-3}$ (red line) and
$5\times 10^{-3}$ (blue line). The pure LMA spectrum is presented by the 
dashed black line, with a normalization factor $f=0.91$ to
reproduce the total observed number of events.
We use the  $^8 B-$neutrino flux
according to the GS98 solar model \cite{GSmodel}. 
}
\label{fig:sk1}
\end{center}
\end{figure}

\begin{figure}[ht]
\begin{center}
\includegraphics[width=13cm]{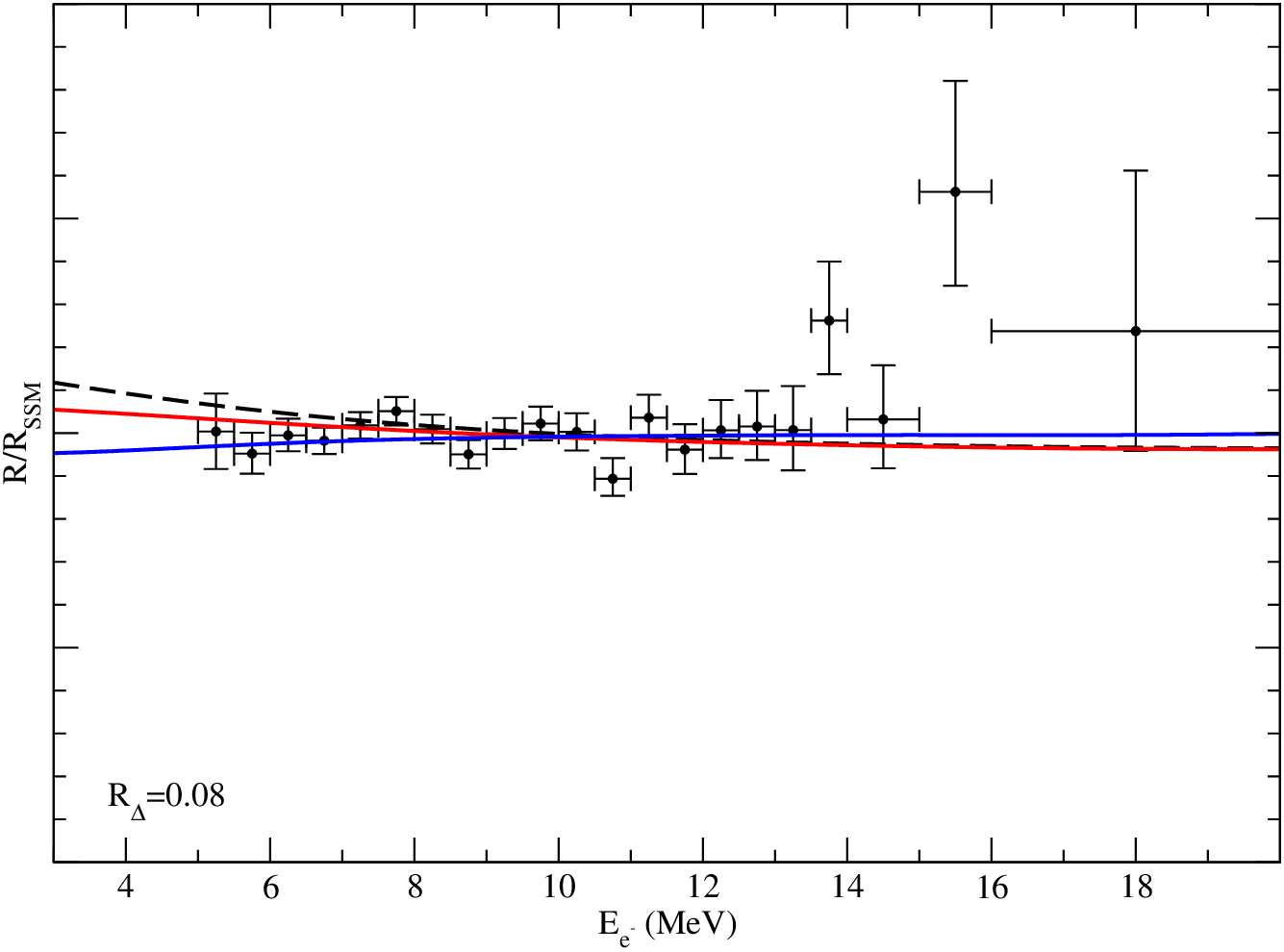}
\caption{The same as in fig.~\ref{fig:sk1} for $R_\Delta=0.08$.
}
\label{fig:sk108}
\end{center}
\end{figure}

\begin{figure}[ht]
\begin{center}
\includegraphics[width=13cm]{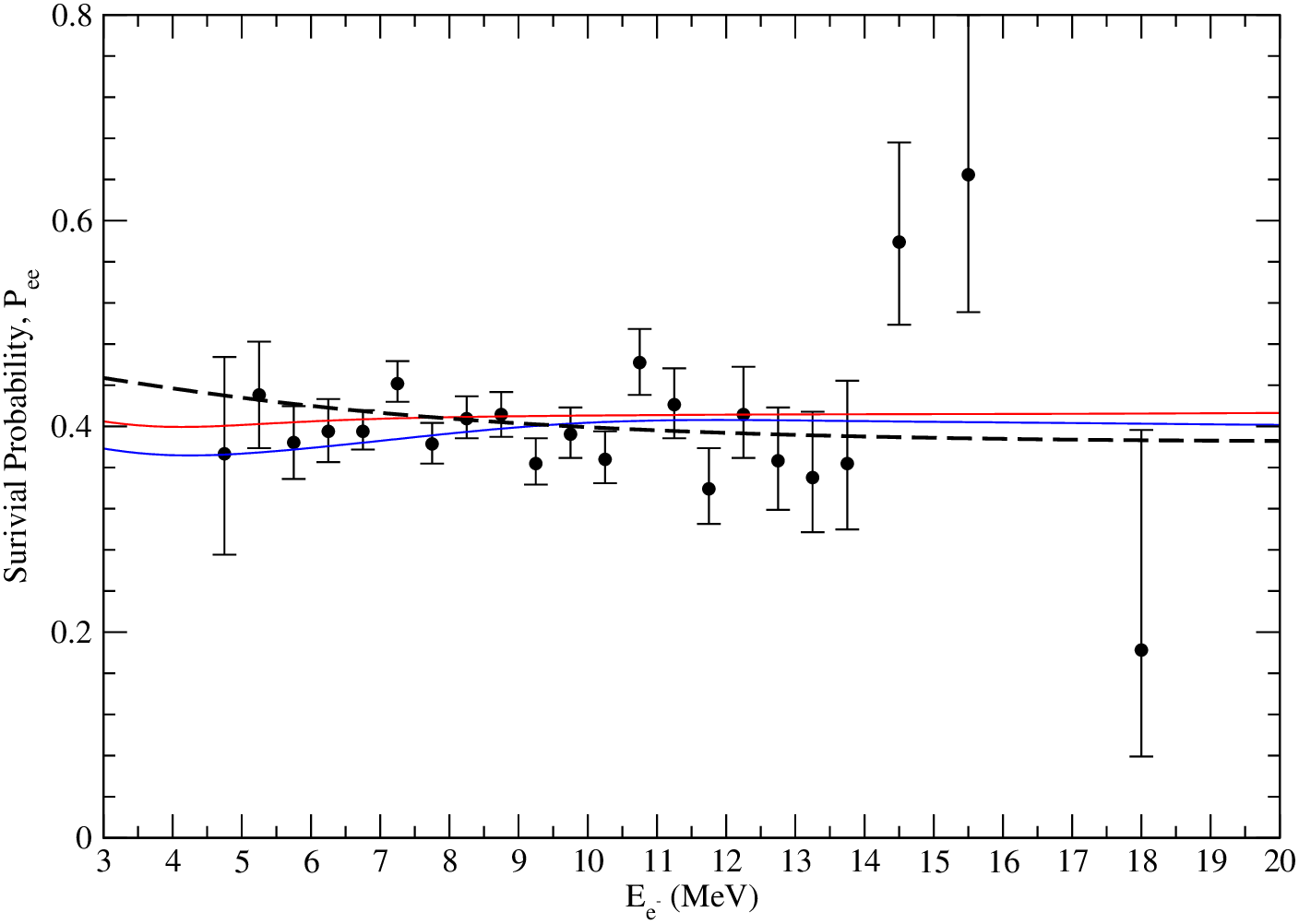}
\caption{Prediction for SuperKamiokande-III
spectrum  versus  experimental data \cite{Abe:2010hy}. 
The neutrino parameters and the solar model as 
well as the normalization factor for pure LMA spectrum are the
same as in fig.~\ref{fig:sk1} (left).}
\label{fig:sk3}
\end{center}
\end{figure}

\begin{figure}[ht]
\begin{center}
\includegraphics[width=13cm]{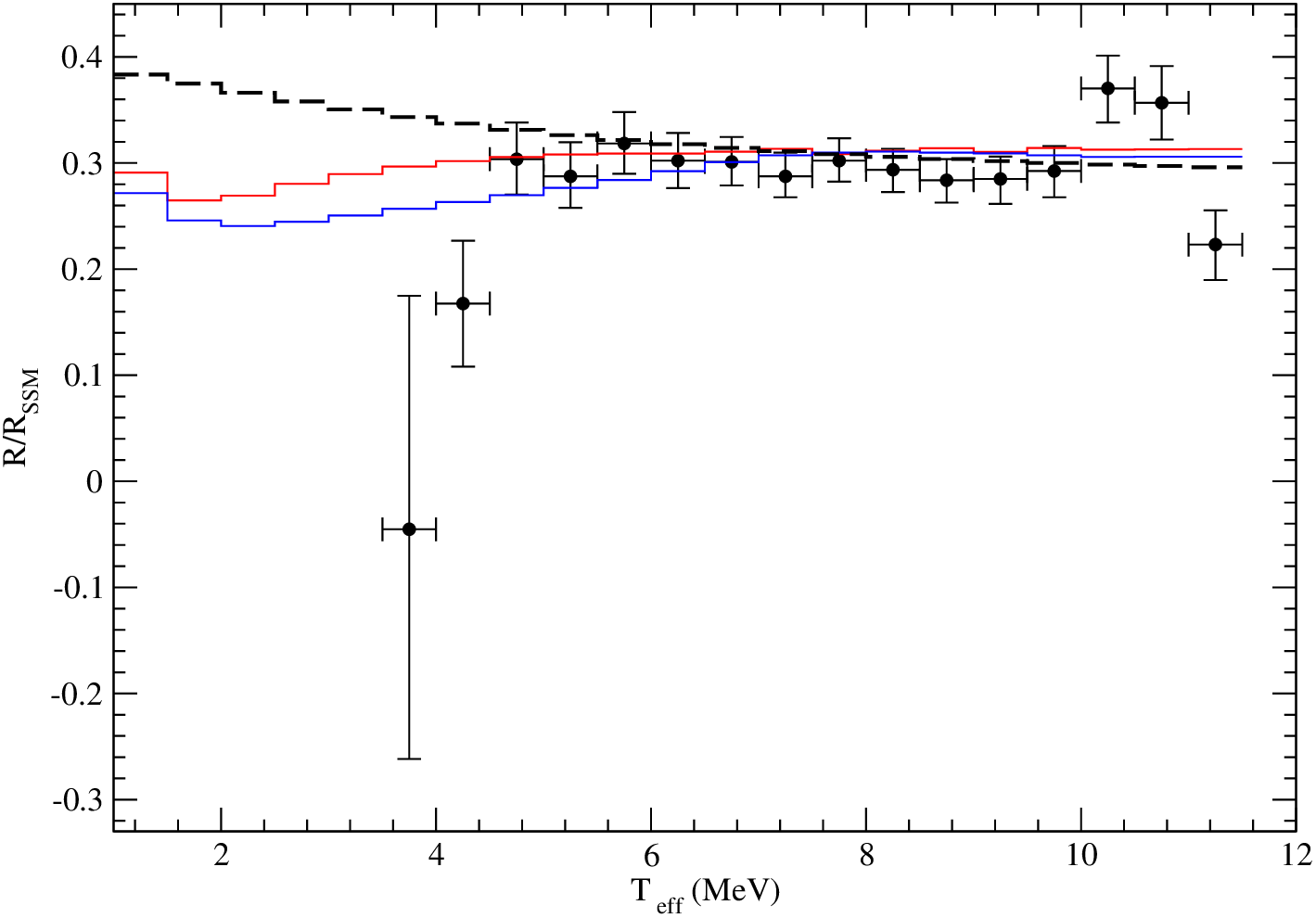}
\caption{Prediction for the SNO-LETA
electron spectrum versus experimental data \cite{Aharmim:2009gd}. 
The neutrino parameters and solar model are the
same as in fig.~\ref{fig:sk1}.
}
\label{fig:snoleta}
\end{center}
\end{figure}

\begin{figure}[ht]
\begin{center}
\includegraphics[width=13cm]{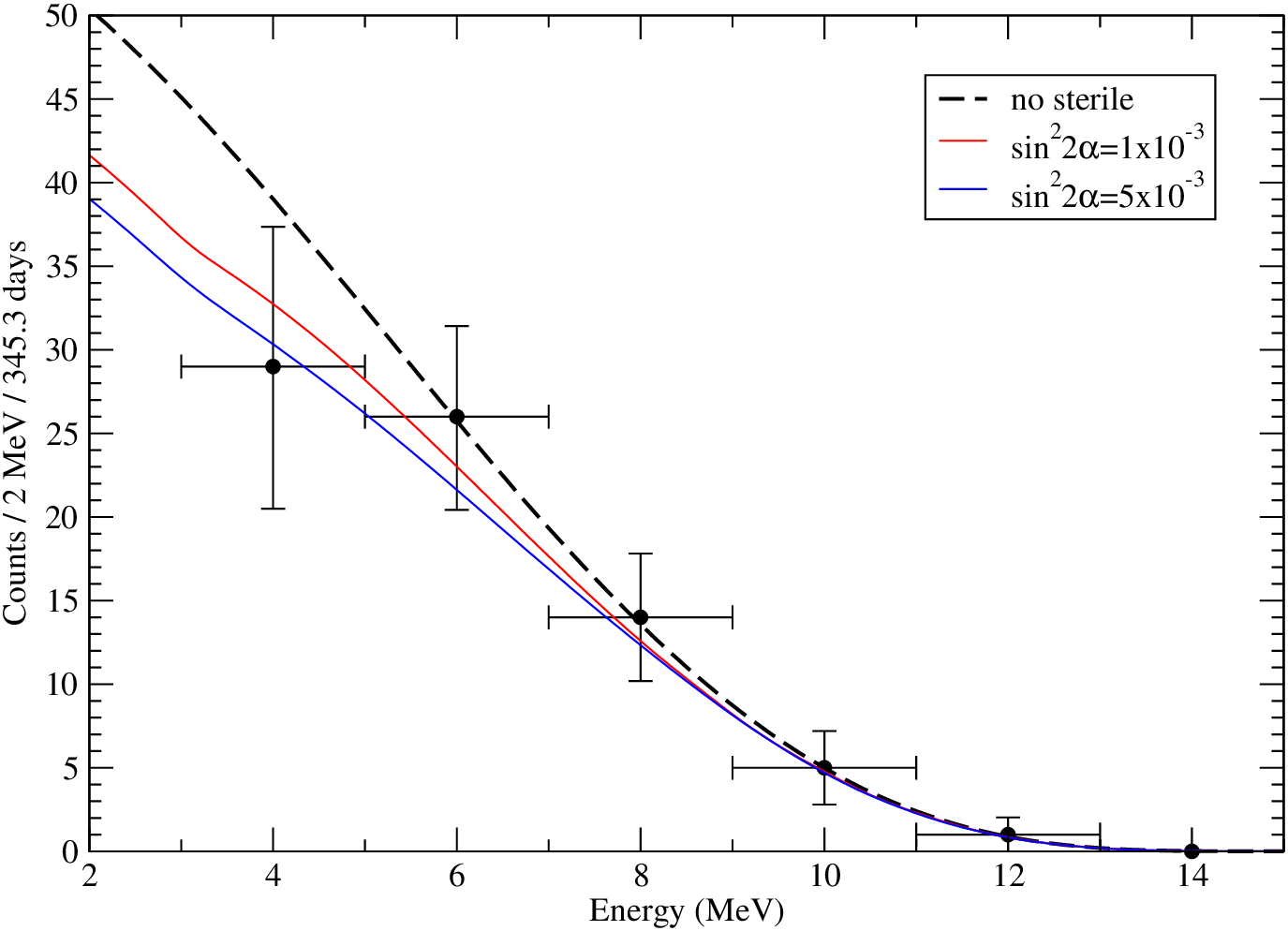}
\caption{Prediction for $B-$neutrino spectrum at Borexino versus  
with experimental data \cite{Bellini:2008mr}. The neutrino parameters and solar model are the
same as in fig.~\ref{fig:sk1}.}
\label{fig:borexino}
\end{center}
\end{figure}

Using the survival probabilities obtained in  
sect. 2 we have computed the energy spectra of events  
for different experiments with and without sterile neutrino. 
These spectra together with experimental data are presented 
in figs.~\ref{fig:sk1} - \ref{fig:borexino}. 
We did not searched for the best fit of the data points,   
and the figures have an illustrative character. 
Notice that due to uncertainty in the original boron neutrino flux 
the experimental points can be shifted with respect to 
the theoretical lines by about $15 \%$.

In figs.~\ref{fig:sk1} and \ref{fig:sk108} we show the ratio of the 
number of events in SuperKamiokande-I (SK-I) 
with and without oscillations for two different values of 
$R_\Delta$. Different curves correspond to the standard LMA solution (dashed) and 
the spectra with conversion to sterile neutrino. 
In the presence of sterile neutrino mixing the upturn can be completely eliminated 
and even transformed into turn down of the spectrum. 
In fig.~\ref{fig:sk1} the dip at 
$E \sim 4$ MeV corresponds to the dip in the probability 
at approximately the same energy 
as in the fig.~\ref{fig:psurv}  (middle pannel). 
The difference of the predictions with and without sterile 
neutrino can be as big as $(15 - 20) \%$ at $E_e < 5$ MeV. 

The SuperKamiokande-III data (SK-III)  (fig.~\ref{fig:sk3}) has  
additional lower energy \cite{Abe:2010hy}, however statistics 
is lower than in SK-I.  Again there is no clear indication 
of the upturn  in the SK-III spectrum
and theoretical lines with sterile neutrino mixing can describe the data better than 
pure LMA solution. 

SNO (fig.~\ref{fig:snoleta}) is more sensitive to distortion of the neutrino spectrum. 
However, the dip in the electron spectrum is shifted to low energies by the threshold 
of the CC reaction on the deuteron: $E = 1.44$ MeV. 
Experimental points are from the SNO-LETA 
charge current event analysis \cite{Aharmim:2009gd}. 
Two low energy points of the spectrum show a sharp  turn down. 
This can not be reproduced  by the proposed dip, although with the dip 
the description is better~\footnote{Too sharp decrease of signal in the 
lowest energy bins is probably statistical fluctuations or some systematics.}. 
Also the Borexino spectrum (fig.~\ref{fig:borexino})  
can be fitted better in presence of sterile neutrino mixing.

According to fig.~\ref{fig:sk1} - \ref{fig:borexino} an improved
 description of the data can be achieved with 
$$
\Delta m_{01}^2 \geq   1.5 \cdot 10^{-5} {\rm eV}^2, ~~~~ \sin^2 2\alpha \sim  10^{-3} .   
$$

\subsection{Further tests}

With higher statistics 
Borexino will improve precision of measurements of the boron 
neutrino spectrum. Also SuperKamiokande  will achieve better 
measurements of spectrum at lower energies. 
The KamLAND solar \cite{Kishimoto:2008zz} and $SNO+$ 
\cite{sno+} experiments will further check the presence of the dip. 

Additional probe of the existence of sterile neutrino  
(and restriction on its  parameters) can be 
provided by measurements of the $pep-$ neutrino line 
with $E = 1.44$ MeV since  the $pep-$neutrino 
flux is known with high precision.  In fig.~\ref{fig:pmonoenergetic} 
we show dependence of the suppression factor for the $pep-$neutrinos as function of $R_\Delta$. 
With increase of $R_\Delta$ the dip shifts to higher energies.  
In the interval  $R_\Delta = 0.07 - 0.11$, the 
$Be-$neutrino flux has the LMA suppression, whereas the $pep-$flux can be suppressed 
by factor 0.15 - 0.20 (the LMA suppression is 0.52). In the range $R_\Delta > 0.12$ 
both fluxes have the LMA suppression. 

In the range $R_\Delta > 0.12$ the $CNO-$neutrinos  are sensitive to the 
presence of the dip, however the original fluxes of these neutrinos are not well 
known. 
 

\section{Extra radiation in the Universe and $\nu_s - \nu_3$ mixing}


Smallness of mixing of the sterile neutrino in the states 
$\nu_{1}$ or/and $\nu_{2}$ ($|U_{si}|^2 < 10^{-3}$) does not lead to production of significant concentration of $\nu_s$ in the Early Universe via neutrino oscillations \cite{pedroS}. 
However, substantial abundance of $\nu_s$ can be generated if $\nu_s$ 
mixes in the state $\nu_3$ and  $U_{s3}$ is large enough. 
Description of the solar neutrino data presented 
in the previous sections 
does not change substantially, if $\nu_s$ mixes with combination 
$$
\nu_\tau'  \equiv  \cos \theta_{23} \nu_\tau + 
\sin \theta_{23} \nu_\mu  \approx \nu_3,  
$$ 
where $\theta_{23}$ is the standard 2 - 3 mixing angle. 
The $\nu_s - \nu_\tau'$ mixing can be parametrized by the angle $\beta$ as
\be
\nu_3 \approx \cos \beta \nu_\tau' + \sin \beta \nu_s, ~~~~
\nu_0 \approx \cos \beta \nu_s - \sin \beta \nu_\tau', 
\label{eq:s0mix}
\ee
so that $U_{s3} \approx \sin \beta$. 
Here we neglect small rotations by the angles $\alpha$ and $\theta_{13}$ 
which do not influence conclusions of this section. 
(These mixings can be introduced before or after the rotation 
(\ref{eq:s0mix})). 
Since $\Delta m_{01}^2 \ll \Delta m_{21}^2 
\ll \Delta m_{31}^2$, the  mass squared difference of 
$\nu_3$ and $\nu_0$ equals 
$$
\Delta m_{30}^2 \approx \Delta m_{31}^2 = 2.5 \cdot 10^{-3}~ {\rm eV}^2. 
$$
For this value of $\Delta m_{30}^2$ the mixing angle 
$\beta$ is restricted by the atmospheric neutrino  
data \cite{Cirelli:2004cz}: 
$$
\sin^2 \beta \leq 0.2 - 0.3, ~~~ (90 \% ~C.L.)   
$$
and by the MINOS searches for depletion of the neutral current events~\cite{Adamson:2010wi}. 
For zero 1-3 mixing the bound  $\beta < 28.8^{\circ}$  has been established 
~\cite{Adamson:2010wi} which corresponds to    
$$
\sin^2 \beta \leq 0.23, ~~~ (90 \% ~C.L).    
$$
In the presence of non-zero 1-3 mixing the bound becomes much weaker. 

If   $\sin^2 \beta \sim 0.2$, then according to \cite{Dolgov:2003sg} 
the sterile neutrinos practically equilibrate before the BBN 
epoch both in the resonance channel 
and in non-resonance channels, {\it i.e.} in neutrino and antineutrino channels. 
Consequently,  in the epoch of nucleosynthesis and 
latter the additional effective number of neutrinos is  
$$ 
\Delta N_{eff} \approx 1. 
$$
The value $\Delta N_{eff} \approx 0.8$ can be obtained for 
$\sin^2 \beta \approx 0.03$ in the non-resonance channel and 
$\sin^2 \beta \sim 10^{-3}$ in the resonance channel. 
According to \cite{Cirelli:2004cz}   $\Delta N_{eff} \approx 0.8$ 
is generated, if the $\nu_s - \nu_\mu$ mixing is about $\sin^2 \beta = 0.02$. 

The CNGS experiment has also some potential to restrict $\sin^2 \beta$ \cite{Donini:2007yf}. 

Let us consider other phenomenological consequences of 
the $\nu_\tau' - \nu_s$ mixing. 
The level crossing scheme can be obtained from fig.~\ref{fig:eigenvalues} 
by adding the third active neutrino level and expanding whole the picture to the left. 
With increase  of density the $\lambda_2$  increase until 
the  the region of 1-3  resonance and then turns down and decreases in parallel to 
$\lambda_1$. Consequently, the sterile level $\lambda_s \approx \lambda_0$  (horizontal line) will cross $\lambda_2$ at some density above the 1-3 
resonance density. 
Thus the mixing of $\nu_s$ in $\nu_3$ leads to appearance 
of the resonance in $\nu_\tau' - \nu_s$ channel (normal mass hierarchy) 
at the density determined by 
$$
V_n = \frac{1}{\sqrt{2}} G_F n_n  \approx \frac{\Delta m_{03}^2}{2E} \approx \frac{\Delta 
m_{31}^2}{2E}.   
$$
In the isotopically neutral medium this density is about 2 time larger 
than the density of 1-3 resonance. 
For the inverted mass hierarchy the resonance appears in the 
antineutrino channel $\bar{\nu}_\tau' - \bar{\nu}_s$. 

Inside the Earth the $\nu_\tau' - \nu_s$ resonance energy equals 
$E \approx 12$ GeV and  wide resonance 
peak appears in the range (10 - 15) GeV. 
This can be tested in studies of the atmospheric neutrinos
(spectra, zenith angle dependences) in 
the IceCube DeepCore detector \cite{deepcore} and in next generation 
Megaton-scale experiments \cite{pedronew}. Effect of such a mixing  
should show up in the long baseline experiments as the energy dependent 
disappearance of the $\nu_\mu-$flux.

The  $\nu_\tau' - \nu_s$ mixing also influences the supernova (SN) 
neutrino conversion.  
The corresponding level crossing in the collapsing star will be 
adiabatic (at least before shock wave arrival) and therefore 
$\nu_\tau'$  converts almost completely in this resonance  
into $\nu_s$. At larger distances from center of a star this $\nu_s-$flux will encounter 
the lower density $\nu_s$ resonances due to the crossing of 
$\nu_s$ and $\nu_{1m}$ levels (see fig.~\ref{fig:eigenvalues}). 
The latter will lead to partial conversion of $\nu_s$ into $\nu_e$,  
since the  adiabaticity is broken in these resonances. 
Hence,  the following chain of transitions is realized: 
\be
\nu_\tau'  (\nu_\mu, \nu_\tau)  \rightarrow \nu_s \rightarrow  \nu_s,  ~ \nu_e. 
\label{eq:sntran}
\ee
Consequently,  even for relatively large 1-3 mixing which leads to 
the  transition 
$\nu_e \rightarrow \nu_3$ with $|\langle \nu_e| \nu_3 \rangle|^2 \ll 1$  
(for normal mass hierarchy), the $\nu_e$ signal may not 
be strongly suppressed due to conversion described in (\ref{eq:sntran}). 
In the case of inverted mass hierarchy similar consideration holds 
for the antineutrino channels. 

In this consideration for simplicity we have neglected possible collective effects 
due to neutrino-neutrino scattering and effects of shock wave propagation 
(see \cite{pedronew}).

\section{Conclusions}

1. Recent measurements of the energy spectra of the solar neutrino events at 
SuperKamiokande, SNO, Borexino do not shown  the expected (according to LMA)
upturns at low energies. The absence of the upturn 
can be explained by mixing of very light sterile neutrino 
in the mass states $\nu_1$ or/and  $\nu_2$ with 
$\Delta m^2_{01} \sim (0.7 - 2) \cdot 10^{-5} $ eV$^2$ ($R_\Delta = 0.07 - 0.25$) and mixing 
$\sin^2 2 \alpha = (1 - 5) \cdot 10^{-3}$. 
Such a mixing leads to appearance of the dip in the 
$\nu_e-$ survival probability 
in the energy range (1 - 7) MeV, thus removing the upturn 
of the spectra. For $\Delta m^2_{01} \sim 2 \cdot 10^{-5} $ eV$^2$ 
and $\sin^2 2 \alpha \sim 5 \cdot 10^{-3}$ 
the $\nu_e - \nu_s$ conversion can even produce 
a turn down of the spectra. Description of the existing solar neutrino data 
in the presence of mixing with sterile neutrino is 
apparently improved. 

Values of $\Delta m^2 < 0.6 \cdot  10^{-5}$ eV$^2$ (for mixing angle interval
$\sin^2 2 \alpha = (1 - 5) \cdot 10^{-3}$)  
are excluded by the Borexino measurements of the $Be-$neutrino flux. 
The presence of the dip can be further tested in future precision measurements 
of the low energy part of the 
$B-$neutrino spectrum as well as the $pep-$ neutrino flux.\\ 

\noindent  
2. Mixing of $\nu_s$ in the $\nu_3$ mass eigenstate with $|U_{s3}|^2 \sim 0.02 - 0.2$ 
leads to production of significant concentration of $\nu_s$ via oscillations in the Early Universe. For $|U_{s3}|^2 \sim 0.1 - 0.2$ nearly equilibrium 
concentration can be obtained both in neutrino and 
antineutrino channels thus generating additional effective 
number of neutrinos $\Delta N_{eff} \sim 1$ before the BBN epoch. 
This can explain recent cosmological observations. \\

\noindent
3. Mixing of $\nu_s$ in $\nu_3$ leads to a number of phenomenological consequences, 
in particular, it can affect the atmospheric and accelerators neutrino fluxes
as well as fluxes of the SN neutrinos. The mixing leads to existence of the  
$\nu_s - \nu_\tau'$ resonance. For neutrinos crossing the Earth 
the resonance should appear at energies $E \sim 10 - 15$ GeV. This can be tested in 
future atmospheric neutrino studies 
with Megaton-scale detectors as well in the long baseline  experiments with 
accelerator neutrino beams.

\section*{Acknowledgments}

P. C. de Holanda is grateful to the AS ICTP for hospitality during 
his visit where most part of this paper has been accomplished. 


\section*{Appendix A: Wiggles}

\begin{figure}[ht]
\begin{center}
\includegraphics[width=10cm]{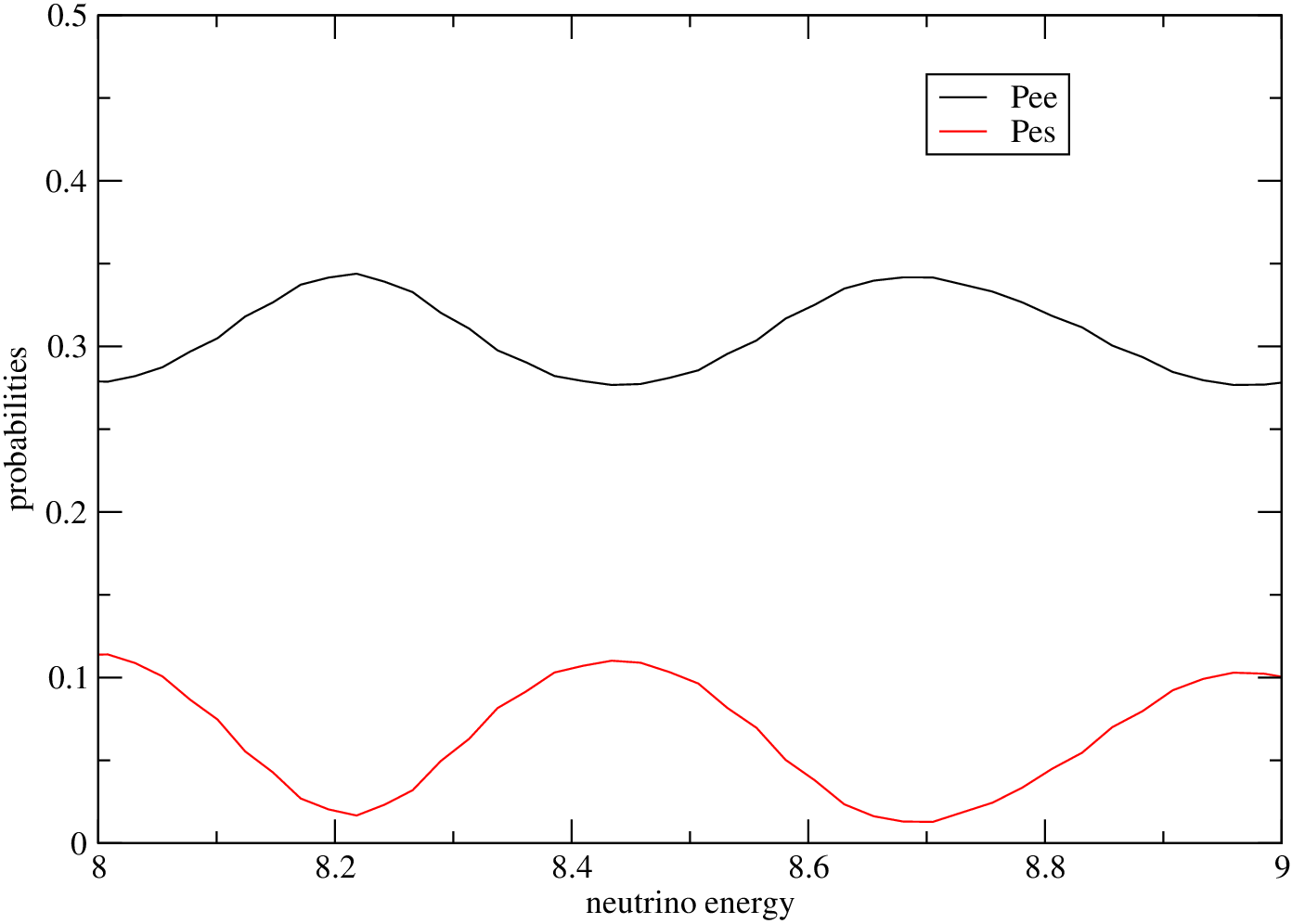}
\caption{A zoomed view of the survival and transition probabilities
in the energy range  where the wiggles can be well seen. 
The sterile neutrino parameters equal $R_\Delta = 0.25$ and $\sin^22\alpha=10^{-3}$.}
\label{fig:probwiggles}
\end{center}
\end{figure}

As we described in sect.~2, wiggles in the dependence of the 
$\nu_e-$survival probability on energy are the result of interference of the amplitudes which contribute to the same $\nu_e \rightarrow \nu_1$ transition. The zoomed 
view of the survival, $P_{ee}$, and transition, $P_{es}$, probabilities 
is shown in fig.~\ref{fig:probwiggles}. The period of wiggles is about (0.5 - 0.6) MeV. 

\vskip 1cm

\begin{figure}[ht]
\begin{center}
\includegraphics[width=13cm]{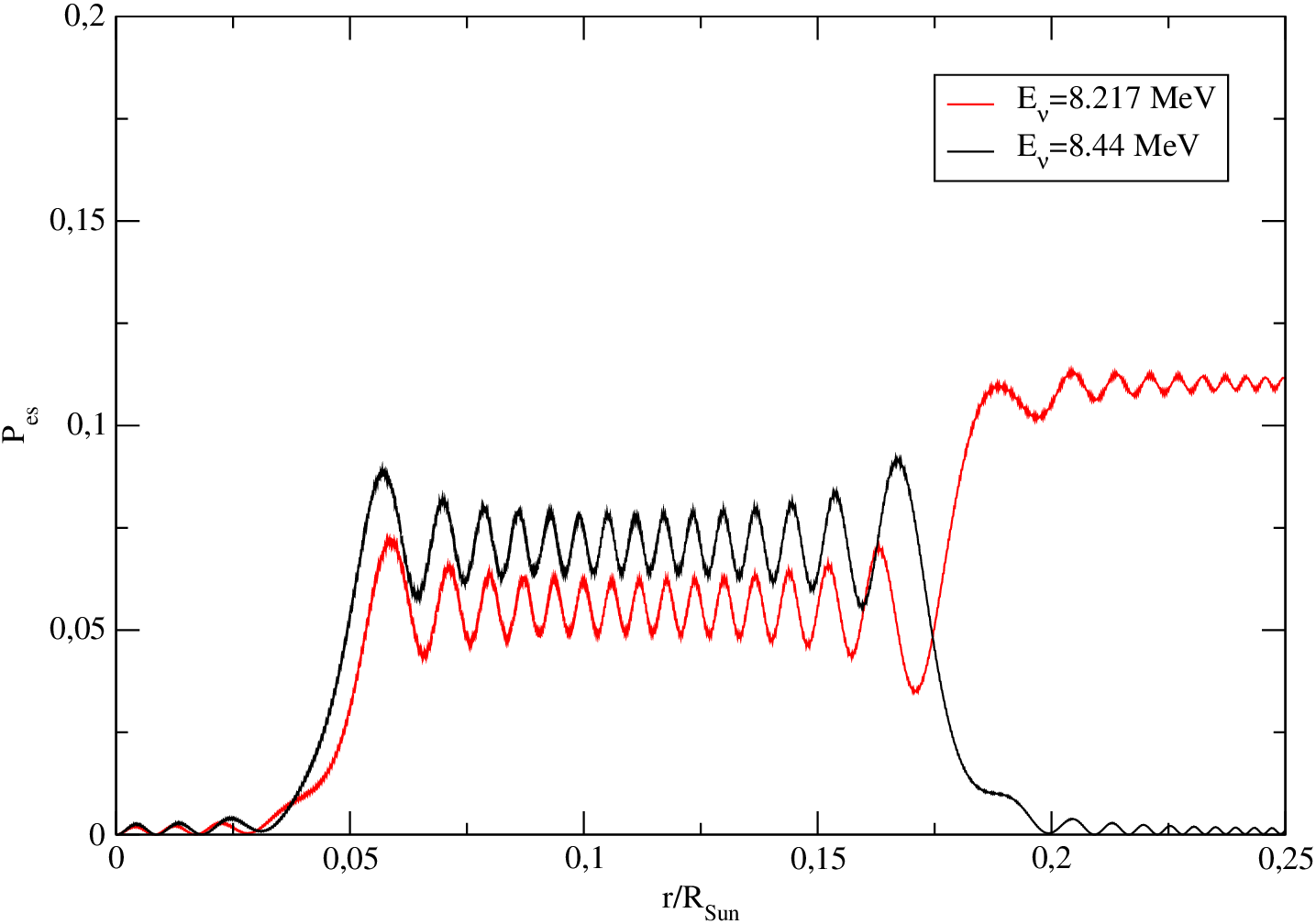}
\caption{Transition probability to sterile neutrino for neutrino
  created at the Sun center as function of distance from the center of the Sun. 
  The sterile neutrino parameters are
  $R_\Delta = 0.25$ and $\sin^22\alpha=10^{-3}$.}
\label{fig:prob_nus}
\end{center}
\end{figure}

The key feature which leads to the wiggles with rather large period 
in the energy scale and therefore prevents them from being averaged out 
at the integration over the production region is that the interference phase is 
collected over relatively small distances $L$. These are 
the distances between the production point 
and the low density $\nu_s-$resonance or the distance between the two 
$\nu_s$ resonances as can be seen in the fig.~\ref{fig:prob_nus}. 
For the neutrino energy $E \sim 8$ MeV the distance $L \approx 20 l_m$ 
where $l_m$ is the  oscillation length in matter. 
Therefore the period of wiggles can be estimated as 
$\Delta E /E \sim l_m/L \sim 1/20$ in agreement with results 
of fig.~\ref{fig:prob_nus}. 

\vskip 1cm

The wiggles are partially averaged 
due to integration over the production region. Notice that with 
decrease of  $\Delta m^2_{01}$ the lower resonance shifts to lower 
densities and the distance $L$  increases leading to smaller period of wiggles and 
stronger averaging. This is one of the reasons of disappearance 
of wiggles with decrease of $\Delta m^2_{01}$. The amplitude of 
wiggles also decreases with increase of $\alpha$: the latter means better adiabaticity and 
therefore suppression of the contribution of one of the channels 
responsible for interference.


\end{document}